\documentclass[useAMS,usenatbib]{mn2e}

\usepackage{lscape}
\usepackage{longtable}

\usepackage{graphicx}

\title[I. Apples to apples $A^2$: photo-z predictions for next-generation surveys]{I. Apples to apples $A^2$: realistic galaxy simulated catalogs and photometric redshift predictions for next-generation surveys}
  
\author[Ascaso, Mei  \& Ben\'itez]{B. Ascaso$^{1}$ \thanks{E-mail:
begona.ascaso@obspm.fr};  S. Mei$^{1,2}$, N. Ben\'itez$^{3}$\\
$^{1}$GEPI, Observatoire de Paris, CNRS, Universit\'e Paris Diderot, 61, Avenue de l'Observatoire 75014, Paris  France\\
$^{2}$University Denis Diderot, 4 rue Thomas Mann, 75205 Paris, France\\
$^{3}$Instituto de Astrof\'isica de Andaluc\'ia (IAA-CSIC), Glorieta de la Astronom\'ia s/n, 18008, Granada, Spain\\}

\begin{document}

\date{Accepted . Received }


\maketitle

\label{firstpage}

\begin{abstract}

We present new mock catalogues for two of the largest stage-IV next-generation surveys in the optical and infrared: LSST and Euclid, based on an N-body simulation+semi-analytical cone with a posterior modification with \texttt{PhotReal}. This technique modifies the original photometry by using an empirical library of spectral templates to make it more realistic. The reliability of the catalogues is confirmed by comparing the obtained color-magnitude relation, the luminosity and mass function and the angular correlation function with those of real data. 

Consistent comparisons between the expected photometric redshifts for different surveys are also provided. Very deep near infrared surveys such as Euclid will provide very good performance ($\Delta z/(1+z) \sim 0.025-0.053$) down to $H\sim24$ AB mag and up to  $z\sim3$ depending on the optical observations available from the ground whereas extremely deep optical surveys such as LSST will obtain an overall lower photometric redshift resolution ($\Delta z/(1+z) \sim 0.045$) down to $i\sim27.5$ AB mag, being considerably improved  ($\Delta z/(1+z) \sim 0.035$) if we restrict the sample down to i$\sim$24 AB mag. Those numbers can be substantially upgraded by selecting a subsample of galaxies with the best quality photometric redshifts. We finally discuss the impact that these surveys will have for the community in terms of photometric redshift legacy. 

This is the first of a series of papers where we set a framework for comparability between mock catalogues and observations with a particular focus on cluster surveys. The Euclid and LSST mocks are made publicly available.

\end{abstract}


\section{Introduction}

In the next decade, several large-area cosmological surveys are expected to be completed in the optical: the Javalambre Physics of the Accelerated Universe Astrophysical Survey (J-PAS, \citealt{benitez09a,benitez14}), expected to start in 2015;   the Dark Energy Spectroscopic Instrument (DESI,  \citealt{levi13}) survey, expecting to start in 2018; the Large Synoptic Sky Telescope (LSST, \citealt{ivezic08,lsst09}) survey, also starting in 2018; the near infrared: the Euclid survey \citep{laureijs11}, which will start in 2020 and the Wide-Field Infrared Survey Telescope (WFIRST\footnote{http://www.ipac.caltech.edu/wfirst/}), which will start in 2020 as well, among others. One of the main goals of such surveys is the measurement of Baryonic Acoustic Oscillations (BAOs) up to redshift 1.3, therefore achieving a combined figure of merit (FoM) superior of 1000, making them to be classified as Stage IV experiments   \citep{albrecht06,albrecht09}. Different strategies have been designed to achieve such goals: J-PAS will be sampling the optical spectrum with 54 narrow-bands, consequently obtaining a superb photometric redshift resolution ($\Delta z <0.003(1+z)$) for all galaxies down to flux limit of $i \sim 23.5$ mag \citep{benitez14};  DESI will target 20-30 million galaxies and quasars spectroscopically over 14000 degrees square; LSST will image a large portion of the sky with 6 broad-band optical bands down to very faint magnitudes ($r \sim 27.5$), consequently obtaining very deep quality data; Euclid will obtain very deep infrared space data ($H \sim 24$ mag), providing a extremely deep and good resolution high-redshift data sample and WFIRST will obtain very deep photometry ($J \sim 27$ mag for 2400 square degrees and $H \sim 29$ mag for 3 square degrees) in six broad bands in the infrared, sampling also the high-redshift universe. All these surveys will be covering thousands of degrees of the northern and southern sky. This, together with the collection of external datasets (X-rays and Sunyaev-Zel'dovich effect maps, spectroscopic samples, etc) will result into an enormous quantity of available data, several order of magnitudes superior to the present data in terms of combined area, depth and photometric redshift resolution.

Up to date, different studies have been performed to estimate the photometric redshift performance of different surveys  \citep{oyaizu08,banerji08,benitez09a,hildebrandt10,bellagamba12,dahlen13,gorecki14,molino14} with a variety of different techniques. Two approaches have been considered in these works. The former uses empirical spectroscopic samples to calibrate the photometric sample while the latter extracts their results from simulations. The first methodology, while using real data, provides usually optimistic results since the spectroscopic samples are usually biased towards brightest samples. Also, these methods provide usually an average global accuracy for the whole population of galaxies at different redshift and magnitude ranges, making these results difficult to be compared consistently.  The second methodology allows a treatment of spectroscopically complete mock samples, hence allowing to provide expected results as a function of magnitude or redshift.

However, it is well known that semi-analytic galaxy formation models are not fully representing the observational universe. Some of the inconsistencies between those models and the observations are related to the inconsistency with the stellar mass function \citep{fontanot09,hansson12,skelton12,weinmann12,henriques12,henriques13,mitchell13}, luminosity function  \citep{fontanot09,cirasuolo10,somerville12,skelton12,hansson12,henriques12,guo13a,gonzalez-perez14,henriques13}, chemical abundances  \citep{fontanot09,delucia12} and colors  \citep{cohn07,weinmann11,skelton12,somerville12,hansson12,henriques12}, consequently producing a wrong tilt of the color-magnitude relation, 'plume' effects of redder galaxies belonging to the cluster spread within 1-2 bright magnitudes, absence of a smooth-transitory green valley and other related effects.  These effects generally lead to an underestimation of the photometric redshift uncertainties. As shown by \cite{arnalte-mur14,ascaso15}, obtaining photometric redshifts directly from mock photometry mimicking the ALHAMBRA survey \citep{moles08} provided a mean photometric redshift dispersion three times higher than the expected for the real data  \citep{molino14}. After  creating a new photometry set using a technique that uses a new empirical library of templates (see \S2.2) to fit the original spectral distribution, the photometric redshift accuracy exactly matched that obtained from real data.

In order to create a solid basis for comparison between different methodologies and datasets, we start with this paper a series of works under the title: \emph{Apples to apples, $A^2$}, which intend to provide a handy well-characterized mock catalogue with realistic well-behaved photometric-redshifts that mimic two of the next generation surveys: LSST and Euclid. A separate work (Ascaso et al. in prep) will also be devoted to analyze the multiple narrow-band J-PAS survey.

In this work, we have applied a similar prescription as in \cite{arnalte-mur14,ascaso15}. The whole point of a applying this prescription is to produce a galaxy sample which resembles as much as possible the real universe in terms of galaxy properties, in particular their distribution of luminosities, redshift, and spectral types. The chosen mock catalogue is based on the publicly available light cone mock catalogues by \cite{merson13}. These catalogs are extracted from an N-body simulation from the Millennium simulation, \citep{springel05} and semi-analytic model of galaxy formation from GALFORM \citep{cole00,bower06}. We have used the  Bayesian Photometric Redshift software 2.0 (\texttt{BPZ2.0}, \citet{benitez00}, Benitez et al 2015, in prep), to obtain photometric redshifts. This technique, called \texttt{PhotReal} (Ben\'itez et al. in prep), is chosen to provide really accurate and realistic results at comparing with real data and avoid fake effects such as 'plume' or unrealistic color-magnitude relations.

At least two more works will continue these  \emph{$A^2$} series. In the second one (Ascaso et al., in prep), we will provide consistent clusters selection functions for all the surveys considered by using the Bayesian Cluster Finder (BCF, \citealt{ascaso12,ascaso14a}), a well behaved detector which is maximizing the advantages of many methods in the literature to detect not only red-sequence cluster and groups. The third $A^2$ paper (Ascaso et al., in prep) will be devoted to forecast the cosmological parameters from the obtained cluster counts in the same consistent way. These mock catalogues will be publicly available and they will be useful for any consistent prediction between the considered surveys: LSST and Euclid in terms of galaxy evolution, galaxy clustering, large-scale structure or any other astrophysical purpose.

The structure of the paper is as follows. In section 2, we introduce the mock catalogue used in this work and we give an overview of the technique developed to provide reliable photometry and realistic photometric redshifts. Section 3, summarizes the main characteristics of the three next-generation surveys considered in terms of area, depth, number of bands and different technical properties. Section 4 provides an extended comparison with different observational dataset to check that their main properties such as stellar masses, magnitudes, colors and spectral types resemble those obtained from the \texttt{PhotReal} post-processed mock catalog. Section 5 describes the obtained results on photometric redshift resolutions and the fraction of outliers for each survey. Section 6 provides a way to improve the overall photometric redshift performance by obtaining the 'best selection samples' with an \emph{odds} quality selection and finally, section 7 summarizes the results of this work and discusses the possible applications of the overall set of data to future studies on providing cluster selection functions and cluster cosmology constraints as well as the impact of this science.

The mock catalogue used in this paper has a fixed cosmology of $\Omega_M=0.25$, $\Omega_{\Lambda}=0.75$, $\Omega_b=0.045$ and $h=0.73$, chosen to match the cosmological parameters estimated from the first year results from the WMAP \citep{spergel03} and which are not significantly different to those obtained by Planck \citep{planck13}. Hence, throughout the whole paper and when necessary, we will use the same cosmology. All the magnitudes in the paper are given in the $AB$ system.

\section{Main simulation}

\subsection{Original light mock catalogue}

In this work, we based our simulations on the original mock catalogs by \cite{merson13}\footnote{http://community.dur.ac.uk/a.i.merson/lightcones.html}. The spatial information of this mock catalogue comes from the Dark Matter halo merger trees extracted from the Millennium Simulation \citep{springel05}, a $2160^3$ particle N-body simulation of the  $\Lambda CDM$ cosmology. This simulation traces the cold dark matter structures from redshift z=127 through the present day in a cubic volume of 500$h^{-1} Mpc^2$. The minimum resolution of this simulation is 20 particles, corresponding to a halo resolution of 1.72 $\times 10^{10}h^{-1}M_{\odot}$.

These haloes have been populated with galaxies created with the variant of the GALFORM  \citep{cole00} semi-analytical model by \cite{bower06}, which models many physical processes governing the formation and evolution of the galaxies such as the star formation and merger history, the radiative cooling of the gas, the feedback as a result of supernovae active galactic nuclei, etc. It also makes predictions for many galaxy properties including luminosity as a function of wavelength. However, these predictions do not always resemble those of the observed galaxies: the colors of the faint galaxies tend to be too red compared to the real ones, there exist a 'plume' of redder galaxies than the red sequence at redshift lower than 1.4, etc. We have overcome these problems by running  \texttt{PhotReal} (see  \S\ref{sec:photoz}), an algorithm that  creates more realistic photometry by using a set of empirical templates to fit the original spectral distribution. Finally, these galaxies have been associated a sub-halo according to its status of central or satellite galaxy at each snapshot. More information on the construction of this mock can be found in \cite{merson13} and references herein.

We have chosen to use the Euclid 500 $deg^2$ wide mock catalogue since it is the deepest of the surveys considered and hence, we can always make it shallower. As for the surveyed area, we chose to use a 500 square degrees mock even if it is a factor of 20 to 40 times smaller than the total area of the targeted surveys as a compromise between having a easy-to-transfer, relatively small in size catalogue and an area large enough to have significant enough statistical results for our scientific purposes.  Indeed, since the main purpose of $A^2$ is to analyze photometric redshift and cluster selection performance, none of these results should be substantially affected after applying a correction which increases the errors following a Poissonian approach to the results to account for this effect. 

As for galaxy clusters and groups, the possible structures that we might be missing due to the limitation of the area would be the rarest ones ($M>10^{15}M_{\odot}$ ) with similar properties as those recently discovered such as, for instance, the SPT-CLJ2344-4243 (\emph{The Phoenix}, \citealt{mcdonald12}), ACT-CL J0102-4915 (\emph{El Gordo}, \citealt{menanteau12}), SPT-CL J0546-5345 \citep{brodwin10}, SPT-CL J2106-5844  \citep{foley11} at redshift $>1$. These structures, which have been reported to be very unfrequent \citep{jee09,hoyle12}, are usually the easiest to detect among other structures at the same redshift almost with any technique (optical, X-rays, cosmic shear) so, we can ensure that if they exist, we will be detecting them. The other possible problem that might arise from using a smaller simulation  would be related with those structures lying on the edges of the cone. However, when detecting galaxy clusters and groups in the next paper or these series, we will use the BCF which, automatically corrects the probabilities by the missing area of the cluster due to edge or masks issues \citep{ascaso12,ascaso14a}.

\subsection{\texttt{PhotReal}: the photometry calibration technique}
\label{sec:photoz}

In order to obtain more realistic and reliable photometry and photometric redshifts, we have applied  \texttt{PhotReal}, a similar algorithm as in previous works \citep{arnalte-mur14,zandivarez14,ascaso15} to the initial mock catalogue. We summarize here this technique developed to  fit the spectral types to make the photometry more realistic. A more extended explanation will be given in a forthcoming work (Ben\'itez et al. 2015, in prep). 

This technique relies on  a library of empirical templates developed for  the well-known photometric redshift code, the Bayesian Photometric Redshift (\texttt{BPZ2.0}, \citealt{benitez00}, Benitez in prep). This library consist of eight empirical templates with a very low outlier rate   and bias in high quality photometric catalogues. Hence, it contains a complete (up to 1-2\%), even if coarse-grained, representation of real galaxy colours for the galaxy populations sampled.
 
The \texttt{PhotReal} procedure starts by running \texttt{BPZ2.0}  in 'ONLY\_TYPE yes' mode  (similarly to  \texttt{BPZ}) to match the optical $ugriz$ and infrared (if available) rest frame photometry of the initial mock catalogue to the best template from the empirical library.  In other words,  \texttt{PhotReal} forces the semi-analytical photometry to have a realistic template distribution by choosing the closest BPZ2.0 template. Once the SED is thus chosen, we generated galaxy fluxes through the survey set of filters  in the same way as in the \cite{benitez09a,benitez09b,benitez14} empirical mocks and add to them empirically calibrated photometric noise. This noise is a combination of the expected photometric noise from the observed relationship between magnitudes and errors in a similar characteristics survey filters (see \S3), plus a systematic noise which is approximately constant with magnitude and most likely unavoidable when measuring galaxy colours in multiband photometry. This systematic is empirically calibrated to be $8\%$ for bluer objects and $6\%$ for red galaxies (Benitez, in preparation).  

Afterwards we run \texttt{BPZ2.0} on those mock catalogues to obtain the final photometric redshift estimation and spectral type. This procedure of course implicitly introduces an error: there may be real galaxies which are present in the catalogues, but not covered by the \texttt{BPZ2.0} templates or not adequately present in the spectroscopic redshift catalogues used to measure the outlier rate. Since we are not aware of any such substantial population within the depths for the surveys considered, we estimate that the procedure followed here does not introduce a significant bias.

The final  \texttt{PhotReal}  output includes photometry and photometric errors, spectral types, stellar masses, a single photometric redshift estimation and the redshift probability function $P(z,T)$, condensing the probability as a function of redshift and spectral type, for all galaxies in the mock.

\section{Considered next generation surveys}

Traditionally, the estimation of the accuracy of photometric redshifts has been given globally, without specifying their dependence with photometry or redshift (although see recent efforts by \citealt{hildebrandt12,dahlen13,raichoor14,molino14}); it has been estimated for different spectroscopic samples with very complicated selection functions to reproduce and different flux cuts have been performed. These issues carry a number of arduousness. First, it becomes complicated to have a clear idea about the best region of the parameters space to use if one wants to select the best quality photometric redshifts. Also, these assumptions make very hard the comparison between the photometric redshift accuracy of different surveys. Furthermore, a clean sample obtained by selecting the best quality photometric redshift has been proved to have many benefits for many scientific purposes but the previous points produce a non equivalent treatment for different surveys.

In the spirit of \emph{$A^2$}, we pretend to perform a systematic comparison between the photometric redshift estimation for different next-generation stage IV surveys using the same mock catalogue and the same photometric redshift technique. In this section, we highlight the main characteristic of the two next-generation surveys we have considered in this work in terms of design, area coverage, wavelength range, photometry depth, complementary data and other technical issues. 

\subsection{LSST}

The Large Synoptic Survey Telescope\footnote{http://www.lsst.org/lsst/}(LSST, \citealt{ivezic08,lsst09}) is a stage IV experiment starting in 2018. The main objectives of this survey are also the constraints of the dark energy parameters, the weak lensing analysis and the study of transients.

The survey is designed to be a large, wide-field ground-based survey imaging the whole visible southern sky from an 8.4m telescope Cerro Pachon (Chile) with six broad-band optical bands $ugrizy$ down to r=27.5 mag after coadding 10 years of operations. The depth of the survey will allow to produce very detailed shear and weak lensing maps in a similar way that was done with the Deep Lens Survey (DLS, \citealt{wittman02,schmidt13,ascaso14a}).The expected depths can be found in Table 1 in \cite{ivezic08}. 

The photometric errors have been estimated according to the description given by the Survey Science Group: eLSST\footnote{http://ssg.astro.washington.edu/elsst/magsfilters.shtml},  e.g. the expected error for a galaxy in a single observation is

\[\sigma_1^2 = \sigma_{sys}^2 + \sigma_{photom}^2\]
with $\sigma_{sys}$ is the systematic photometric error ranging between 0.005-0.01 mag for the LSST case, and $\sigma_{photom}$ is the random photometric error, which can be written as

\[\sigma_{photom}^2 = (0.04-\gamma)x + \gamma x^2,\quad \rm with \quad x = 10^{0.4(m-m_5)} \]
where $m_5$ is the 5$\sigma$ limiting magnitude depth and $\gamma$ depends on the sky brightness and readout noise. Eventually, the final error obtained for $N$ observations is $\sigma_N=\sigma_1/\sqrt{N}$. 

\subsection{Euclid}

The Euclid survey\footnote{http://www.euclid-ec.org/}\citep{laureijs11} is an European space mission survey starting in 2020. Euclid will also be a Stage IV experiment and it is expected to provide very accurate measurements of the dark energy constrains from a number of different probes including BAOs, weak lensing, cluster counts and high-z supernova. 

The survey will have a 'Euclid Wide Survey' covering 15,000 square degrees of the sky in the near infrared ($YJH$), down to $\sim$24 mag in $H$ band providing also near-infrared spectroscopy and two 'Euclid Deep Fields', about 2 magnitude deeper than the wide survey which will cover around 20 square degrees each.  The deep infrared imaging will map and systematically explore the  mostly unknown-high redshift  Universe. 

Euclid space observations will be combined with other space and ground-based observations to obtain the source photometric redshifts and physical properties. Among the optical surveys that will be available form the ground, one is already available, the Sloan Digital Sky Survey (SSDS), and two are planned, the LSST and the DES \citep{des05} surveys. Hence, we will consider two cases which will provide us a maximum and minimum limit in the photometric redshift and cluster survey predictions for the Euclid survey:

\begin{itemize}
\item Euclid pessimistic case: where the optical data counterpart will come from the five broad band photometry $grizY$ from the DES down to the depth stated in Table 1 by \cite{mohr12}. The photometric errors for the DES bands have been estimated from the mock catalogues by \cite{chang14} which mimic the properties of the DES data.
\item Euclid optimistic case: where the optical counterpart will come from the combination of DES and LSST-like surveys.
\end{itemize}

In both cases the photometric errors for the 'Euclid Wide survey' have been calibrated from the Cosmic Assembly Near-infrared Deep Extragalactic Legacy Survey (CANDELS\footnote{http://candels.ucolick.org/},  \citealt{guo13b}) and normalized to the Euclid depth (e.g. H$\sim$24 at 5$\sigma$, \citealt{laureijs11}).

\section{Performance of the mock galaxies}

\subsection{Model properties of the galaxies}
 
With the double-fold objective of measuring the variation of the modeled properties of the galaxies used in this simulation and proving the non-dependence of the results shown here on the photometric redshift code used, we have used \texttt{LePhare} \citep{ilbert06,ilbert09} to obtain an estimation of the stellar masses, ages and star formation rates (SFRs) of the galaxies in the original mock and the \texttt{PhotReal} post-processed  catalogues. 

To do so, we have used 27 stellar population synthesis models generated using \cite{BC03} models at 14 ages that range from 0.1 to 13.0 Gyr. These BC03 models use a \cite{chabrier03} initial mass function with metallicities of Z=0.004, 0.008, and 0.2$Z_{\odot}$ and are characterized by exponentially declining star formation histories (SFHs) with timescales of $\tau=0.1-30$ Gyr. We adopt the \cite{calzetti00} reddening law with $0\le E(B-V)\le0.5$. We first obtained the photometric redshift estimation and we later fixed it to provide model measurement of the galaxies.

\begin{figure}
\centering
\includegraphics[clip,angle=0,width=1.0\hsize]{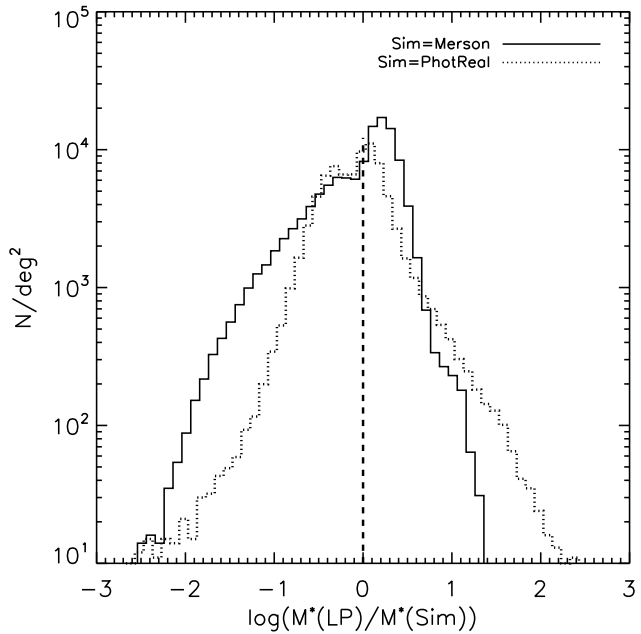} 
\caption{Solid line: Distribution of the difference between the original stellar masses provided in the original mock catalogue and the stellar masses fitted with \texttt{LePhare} to the original mock photometry. Dotted line: Distribution of the difference between the  stellar masses obtained from the \texttt{PhotReal} post-processed mock catalogue and the stellar masses fitted with \texttt{LePhare} to the \texttt{PhotReal} mock photometry. The vertical dashed line indicates the zero value for reference.}
\label{fig:SMdist}
\end{figure}

Then, we have compared on one hand, the \texttt{LePhare} stellar masses estimated from the mock initial photometry with the original stellar masses in the mock and on the other, the \texttt{LePhare} stellar masses estimated from \texttt{PhotReal} post-processed photometry with the stellar masses obtained with \texttt{BPZ2.0}. In Fig. \ref{fig:SMdist}, we display the difference between the \texttt{LePhare} estimation to the original photometry and the initial mock catalogue masses (solid line); and the difference between the \texttt{LePhare} estimation to the \texttt{PhotReal} photometry and the stellar masses estimated with \texttt{BPZ2.0} to this photometry (dashed line). The difference between the initial catalogue masses and \texttt{LePhare} estimations is a factor of two more biased towards negative values ($<SM(LePhare)_i-SM(Merson)>=-0.028 \pm 0.494$) than the difference between the \texttt{BPZ2.0} masses and their estimation with \texttt{LePhare} $(<SM(LePhare)_f-SM(BPZ2.0)>=-0.018 \pm 0.432$). However, the dispersion obtained in the relation is similar.

\begin{figure}
\centering
\includegraphics[clip,angle=0,width=1.0\hsize]{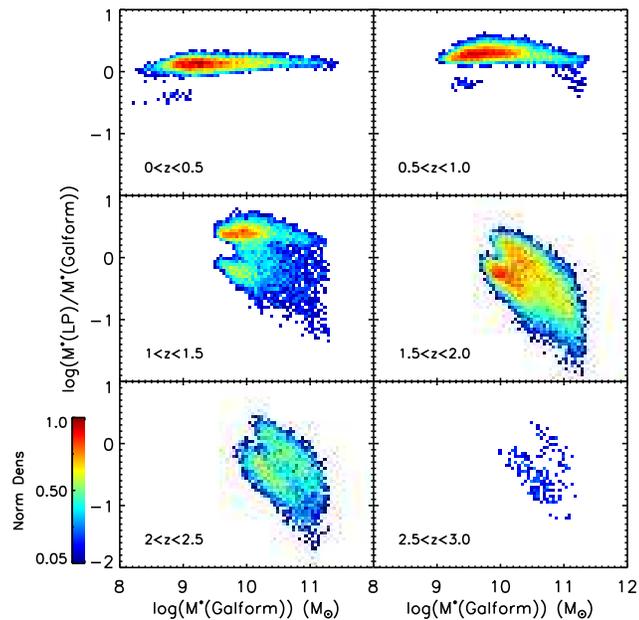} 
\caption{Density map of the logarithmic ratio between the stellar mass given in initial mock catalogue and its fit with  \texttt{LePhare} as a function of the initial mock stellar mass for six different redshift bins.}
\label{fig:SMmassini}
\end{figure}

Looking into this issue more in detail, we have plotted in Fig. \ref{fig:SMmassini}  the density map of the logarithmic ratio between the stellar mass given in initial mock catalogue and its fit with  \texttt{LePhare} as a function of the initial stellar mass for different redshift bins.  While the agreement of the stellar masses at lower redshifts is tight  (0.114), the dispersion increases as we increase in redshift, finding that the masses estimated with \texttt{LePhare}  are systematically smaller than the original ones. A similar behavior has been observed in \cite{mitchell13} when comparing the initial input masses with modeled masses and it has been associated to the presence of dust in the model. 

\begin{figure}
\centering
\includegraphics[clip,angle=0,width=1.0\hsize]{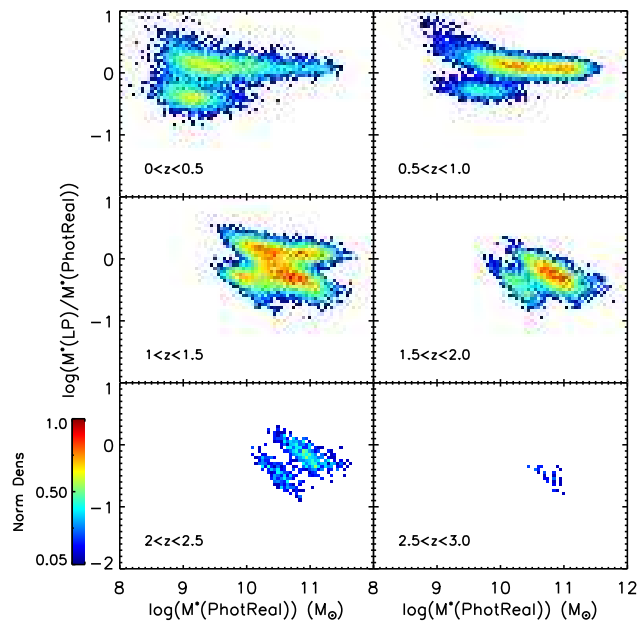} 
\caption{Density map of the logarithmic ratio between the stellar mass obtained from the fit to the  \texttt{PhotReal} photometry and the estimation with  \texttt{LePhare} as a function of the  \texttt{PhotReal} mass for six different redshift bins.}
\label{fig:SMmassfin}
\end{figure}

Similarly, we have plotted in Fig. \ref{fig:SMmassfin}, the logarithmic ratio between the stellar mass obtained from the fit to the  \texttt{PhotReal} photometry and the estimation with  \texttt{LePhare} as a function of the  \texttt{PhotReal} mass as a function of redshift. In this comparison,  we see that in the lower redshift bin, the bias of the stellar mass estimation is smaller than the previous comparison (-0.002). However, the dispersion is 3 times larger. As we increase in redshift, we notice a bimodal distribution of the difference, explained in  \cite{mitchell13} for the use of different Initial Mass Function (IMF) and Stellar Population Synthesis (SPS) models. However, the dispersion do not increase,  becoming comparable between $0.5<z<1.0$ and a factor of 1.5-2 smaller at z$>$1.0.

As a consequence, the \texttt{PhotReal} post-processed catalogue seems to be a factor of $\sim$ 1.5-3 less sensible to the presence of dust in the model. In  \S\ref{sec:smasssec}, we will investigate more in detail the stellar mass function.

\begin{figure}
\centering
\includegraphics[clip,angle=0,width=1.0\hsize]{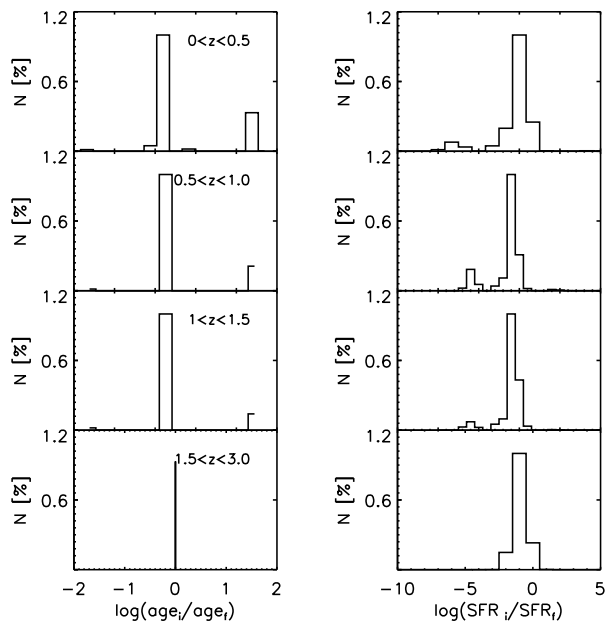} 
\caption{Normalized distribution of the difference in logarithm between galaxy ages (left column) and star forming rates (right column) obtained with \texttt{LePhare} from the initial photometry and the \texttt{PhotReal} post-processed photometry. From top to bottom, the differences are shown as a function of increasing redshift: $0<z<0.5$, $0.5<z<1.0$, $1.0<z<1.5$ and $1.5<z<3.0$, respectively.}
\label{fig:SFH}
\end{figure}

In Fig. \ref{fig:SFH}, we show the differences between stellar ages and star forming rates (SFR)  obtained from the initial and transformed mock catalogue for different redshift bins. The stellar ages of the considered galaxies are very similar, particularly at high redshifts, where they fully agree. Overall, we find that more than $\sim$ 60\% of the sample has a difference in galaxy ages less than 0.5 dex.

As for the SFR, we find that the differences are smaller in the highest redshift bin ($1.5<z<3$), obtaining $<log(SFR_i/SFR_f)>=-0.281 \pm 0.509$. At lower redshift the dispersion increases a factor of 3-4 due to the presence of a secondary peak at all redshift bins with a systematic difference of $log(SFR) <-3$. These secondary peaks indicate that a substantial part of the galaxies in the \texttt{PhotReal} post-processed mock are better fitted by a spectrum with different star forming rates than originally defined, particularly at lower redshifts. This is compatible with the shift to bluer colors shown in the previous section where we show how the \texttt{PhotReal} post-processed catalogue makes the color-magnitude relation (CMR) more realistic showing a blue cloud as expected in observational data.

While the objective of this test is not to obtain reliably model parameter of the galaxies in the mock but rather check the difference between their properties, we conclude that the main star formation history of the galaxies in the mock are not changing substantially, particularly at high redshift.

\subsection{Spectral type recovery}

\begin{figure}
\centering
\includegraphics[clip,angle=0,width=1.0\hsize]{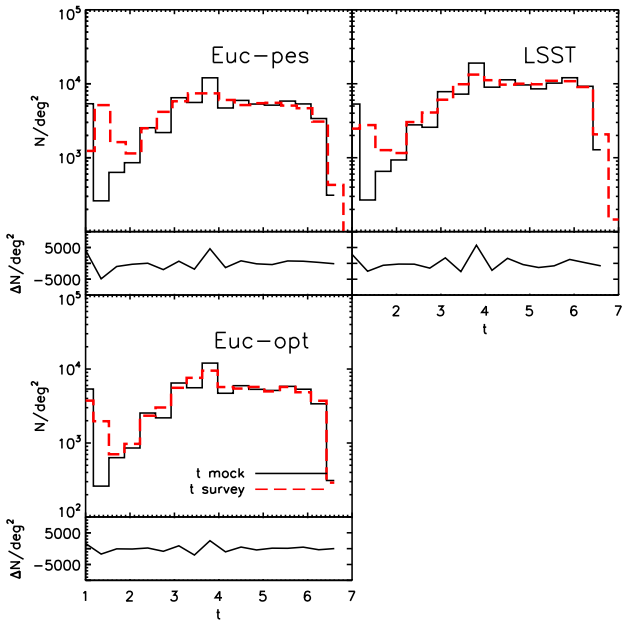} 
\caption{Spectral type distribution obtained from the original Merson photometry ($t_i$, solid line) and from the \texttt{PhotReal} post-processed mock catalogue ($t_f$, red dashed line) for each of the considered surveys: Euclid-Pes (top left panel) and Euclid-Opt (bottom left panel) and LSST (top right panel).  The bottom panel shows the difference between the respective distributions. The Euc-opt shows an excellent recovery of the spectral type, whereas the difference between the original and the recovered simulation for the Euclid-pes and the LSST becomes a factor of 1.5-2 larger.}
\label{fig:types}
\end{figure}

The original mock catalogue by \cite{merson13} did not include the information regarding the spectral type of the galaxy or its spectrum due to disk space storage. We have instead performed a comparison of the spectral type obtained from the fit to the original photometry, $t_i$, and  the recovery spectral type obtained from the \texttt{PhotReal} photometry, $t_f$. The spectral types are similar to the ones used in \cite{benitez00}; the first time refers to a typical Elliptical, type 2 to a Lenticular, type 3-4 to Early and late spirals respectively, and type 5-8 to starburst. In Fig. \ref{fig:types}, we show these type distributions for the Euclid-Pes (top left panel), Euclid-Opt (bottom left panel) and LSST (top right panel) surveys respectively.

The difference between the original and recovered spectral type distribution is minimum for the Euclid-opt case (1.90\%), as expected and it changes slightly for the Euclid-pes case (3.65\%) and the LSST case (2.90\%).  While a higher departure of the final spectral distribution is observed for the LSST and Euclid-pes, we conclude that the spectral distribution is well-recovered for the three cases considered.

\subsection{Color-magnitude relation}

In the original semi-analytical mock catalogue, the first effect that we noticed is the fact that the CMR usually exhibited a lack of blue cloud that it is not observed with observational data (e.g. \citealt{ascaso08,mei09,ascaso14b})  in agreement with  \cite{merson15}.  \texttt{PhotReal} overcomes this problem by  re-computing the photometry and therefore, shifting some galaxies from the red sequence to the blue cloud. Note that the magnitudes that we recover have already incorporated photometric error estimated for the survey whereas the original mock catalogues do not. 

\begin{figure*}
\centering
\includegraphics[clip,angle=0,width=1.0\hsize]{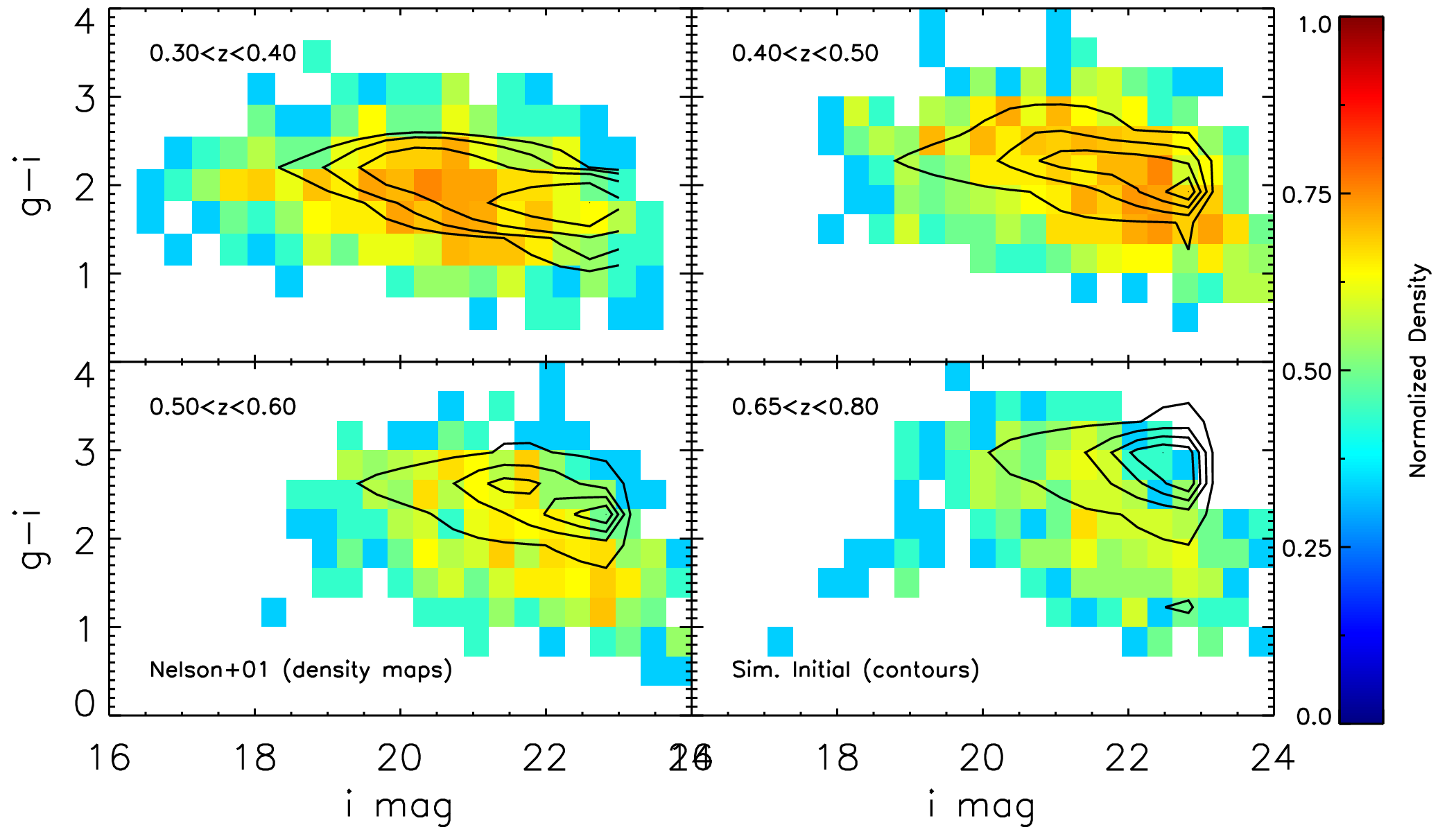} 
\includegraphics[clip,angle=0,width=1.0\hsize]{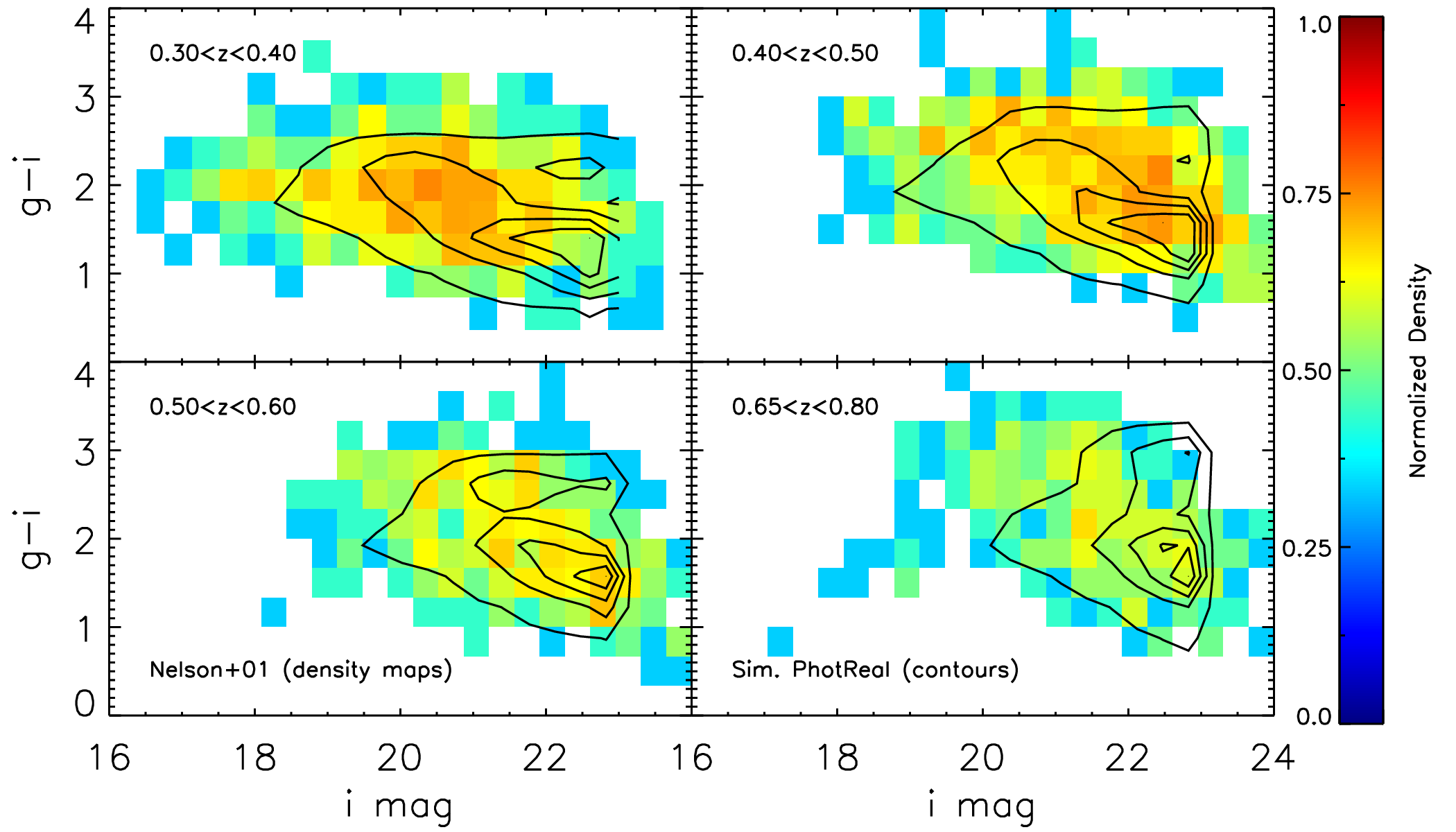} 
\caption{Stacked $g-i$ color versus $i$ magnitude density maps from Nelson et al. 2001 for different redshift bins (0.30$<z<$0.40 top left panel (585 galaxies), 0.40$<z<$0.50 top right panel (713 galaxies), 0.50$<z<$0.60 bottom left panel (430 galaxies), 0.65$<z<$0.80 bottom right panel (296 galaxies) respectively for the four top and bottom panels). The color scale is logarithmic and redder colors represents higher densities while bluer colors refer to less dens regions. We overplottted the density contours of the stacked CMR obtained from the mock photometry using the DES and LSST bands for the same redshift slices. The four top panels display the contours obtained from the initial photometry, whereas the four bottom panels refer the contours obtained from the post-processed \texttt{PhotReal} photometry.  \texttt{PhotReal} reproduces in an accurate way the galaxy color distribution in both the red sequence and the blue valley within the redshift range 0.30$<z<$0.80.}\label{fig:cmr1}
\end{figure*}

In order to illustrate this effect, we displayed in Fig. \ref{fig:cmr1}, the $g-i$ color versus $i$ magnitude density plots resulting of stacking the \cite{nelson01} CMR in redshift bins of 0.1. The CMR information has been extracted using \texttt{Dexter}\footnote{http://dexter.sourceforge.net/} and the original $V-I$ versus $I$ relation have been converted to $g-i$ versus $i$ magnitude by adding a color term. On top of these density maps, we overlapped, in the four top panels, the CMR contours obtained from stacking the original (four top panels) and \texttt{PhotReal}  (four bottom panels) post-processed photometry  using the LSST bands for the same redshift bins.  In all cases, the contours limit are shown to 5\% of the total density CMRs.

\begin{figure*}
\centering
\includegraphics[clip,angle=0,width=1.0\hsize]{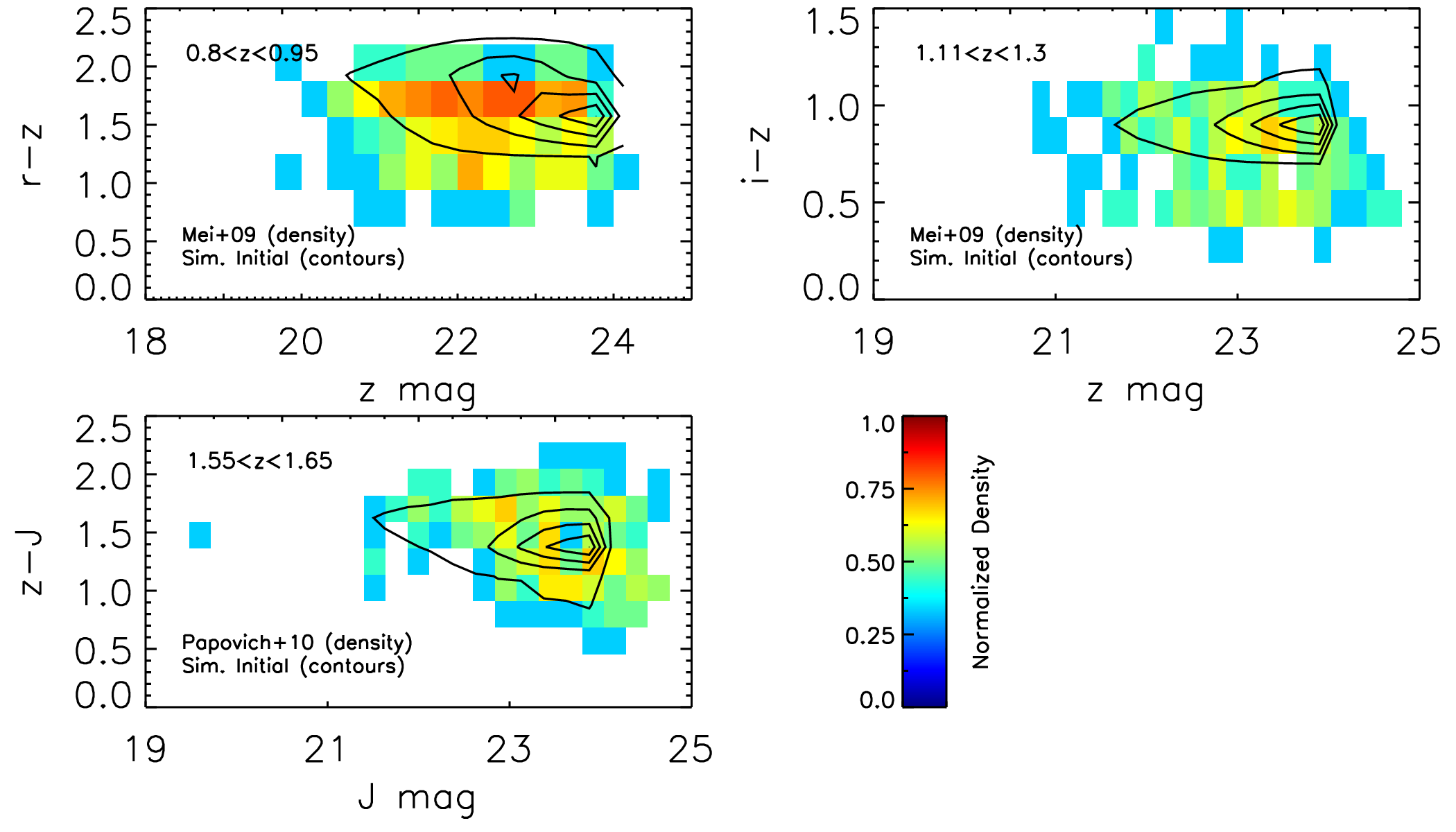} 
\includegraphics[clip,angle=0,width=1.0\hsize]{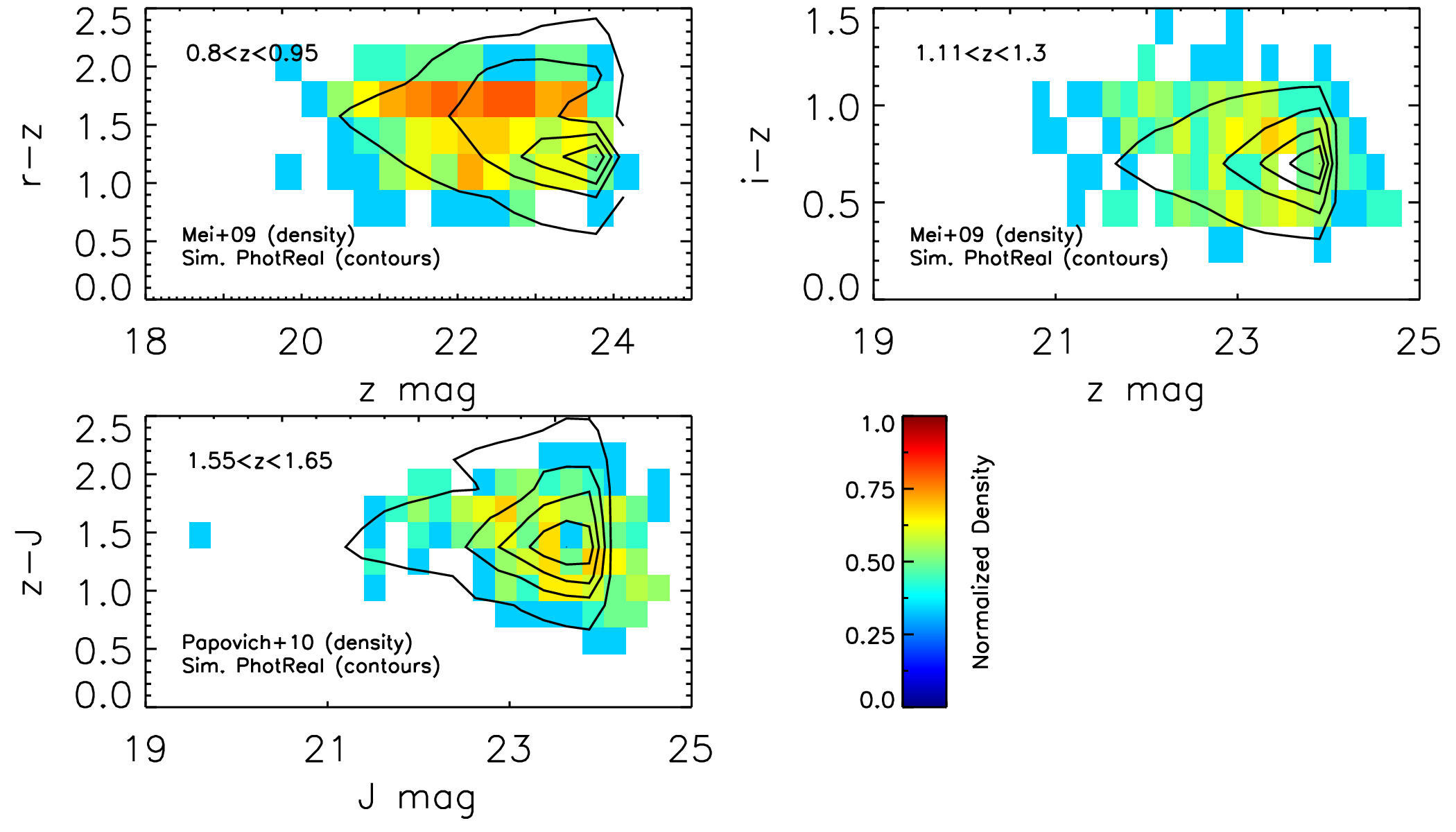} 
\caption{Stacked $r-z$ color versus $z$ magnitude density map for the redshift bin 0.8$<z<$0.95 (332 galaxies); $i-z$ color versus $z$ magnitude density map for the redshift bin 1.1$<z<$1.3 (224 galaxies), both redshift bins from Mei et al. 2009  and  $z-J$ color versus $J$ magnitude for the redshift bin 1.55$<z<$1.65 (266 galaxies) from Papovich et al. 2010 (left top, right top and left bottom panels for the three top and bottom panels respectively). The color scale is the same as in Fig. \ref{fig:cmr1}. We overplotted the density contours of the stacked CMR obtained from the mock photometry using the DES, LSST and Euclid bands for the same redshift slices. The three top panels display the contours obtained from the initial photometry, whereas the three bottom panels refer the contours obtained from the post-processed \texttt{PhotReal} photometry. \texttt{PhotReal} reproduces in an accurate way the galaxy color distribution in both the red sequence and the blue valley within the redshift range 0.80$<z<$1.65.}
\label{fig:cmr2}
\end{figure*}

Similarly, we plotted the CMR density maps at redshift $z> 0.8$ in Fig \ref{fig:cmr2}. The top left panel displays the $r-z$ versus $z$ magnitude at the redshift slice 0.8$<z<$0.95; the top right panel shows the $i-z$ versus $z$ magnitude for the redshift slice $1.1<z<1.3$ and finally, the bottom left panel refers to the $z-J$ versus $J$ magnitude for the redshift bin 1.55$<z<$1.65. The first two CMRs are extracted from \cite{mei09}, also with \texttt{Dexter}, whereas the last CMR corresponds to the galaxy clusters at redshift 1.62 by \cite{papovich10}. These colors are used since they straddle the 4000 \AA break in each redshift range. At the view of these two figures, we observe that \texttt{PhotReal} reproduces the colors of galaxies in the blue valley, in agreement with the observational data, whereas the original mock catalogue displays a lack of blue galaxies.

\begin{figure}
\centering
\includegraphics[clip,angle=0,width=1.0\hsize]{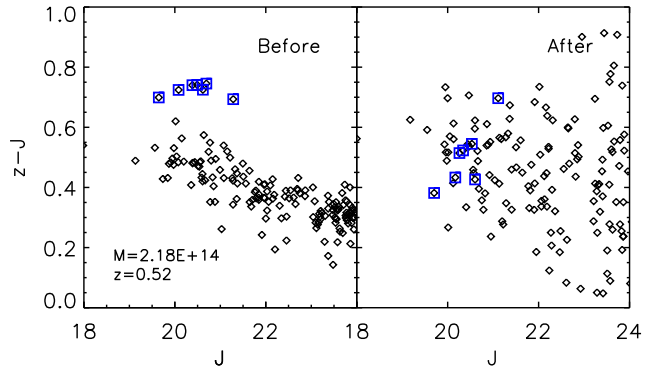} 
\caption{$z-J$ color versus $J$ magnitude diagram for a random case of a  $2.17\times10^{14}M$ cluster at $z=0.52$ in the mock catalogue as seen by the Euclid-opt survey with its bands. The left panel refers to the original mock CMR whereas the right panel represents the recovered CMR after introducing the real effects of the catalogs. The blue squares show the 'plume' galaxies in the original CMR and their location in the \texttt{PhotReal} post-processed CMR.}
\label{fig:plume}
\end{figure}

A secondary effect related to colors, noticed in the original mock catalogues is the presence of a 'plume' of redder galaxies than the CMR. While this effect has been noticed in the infrared colors at high-redshift (z$\sim$1.4), the effect continues appearing at lower redshift in the optical bands. In Fig \ref{fig:plume}, we show an example of the $z-J$ color-magnitude relation of a cluster at $z\sim 0.52$ and $M\sim2.17\times10^{14}M_{\odot}$ as seen by the Euclid-Opt survey in the original mock catalogue (left panel) and in the \texttt{PhotReal} post-processed one (right panel). The blue squares indicate the galaxies considered to be in the 'plume' in the original mock catalogue and their position in the \texttt{PhotReal}  post-processed mock catalogue. We observe that the presence of this systematic 'plume' disappears from the color-magnitude diagrams after the correction but they are still red galaxies, located in the main red sequence of the cluster. These galaxies have not changed substantially their colors and, in addition, they have also diluted their photometry within the photometric errors of the simulation. This result ensures us that, while we are making the mock catalogue more realistic, we are not changing substantially the properties of their galaxies.

\subsection{Luminosity Function}

\begin{figure}
\centering
\includegraphics[clip,angle=0,width=1.0\hsize]{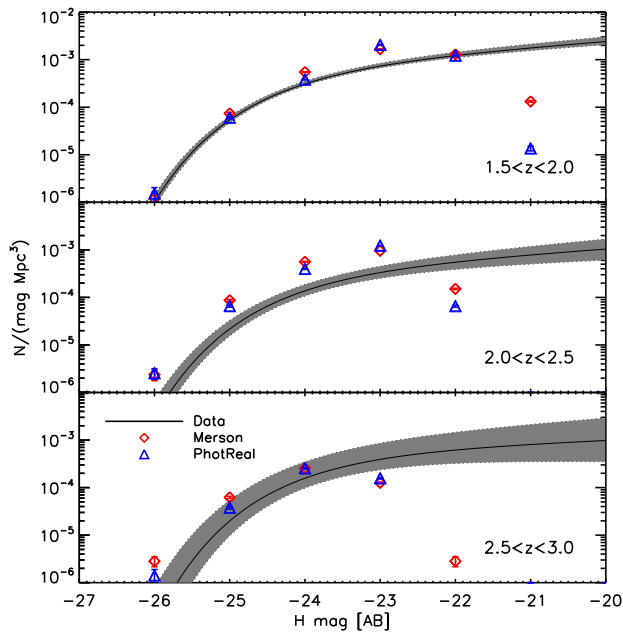} 
\caption{H-band rest-frame luminosity function for three different redshift bins: $1.5<z<2.0$ (top panel), $2.0<z<2.5$ (middle panel) and $2.5<z<3.0$ (bottom panel). The red diamonds refer to the magnitude counts of the initial mock catalogue and the blue triangles refer to the same counts for the \texttt{PhotReal} post-processed mock catalogue for the Euclid-opt survey. The solid line and the gray area are the observational luminosity function obtained by Stefanon \& Marchesini, 2013.}
\label{fig:LFeuc}
\end{figure}

An additional check that we have performed is the comparison between the infrared and optical luminosity function (LF) at all redshift ranges. In Fig. \ref{fig:LFeuc}, we plot the observational H-band rest-frame LF for three different redshift bins at $z>1.5$ obtained from the fits by \cite{stefanon13} from the Multiwavelength Survey by Yale-Chile (MUSYC, \citealt{gawiser06}), the Faint Infrared Extragalactic Survey (FIRES, \citealt{franx00}) and the GOODS Chandra Deep Field-South Survey (FIREWORKS, \citealt{wuyts08}). We overplot the luminosity counts obtained for the Euclid-opt simulation in the original mock catalogue and in the \texttt{PhotReal} post-processed one. We find a very good agreement between both counts from the simulation and the observational LF at least down to $M_H<-21$ where the flux limit of the simulation is being reached. This becomes a further indication that the photometry in the H-band at least it is not significantly altered.

\begin{figure}
\centering
\includegraphics[clip,angle=0,width=1.0\hsize]{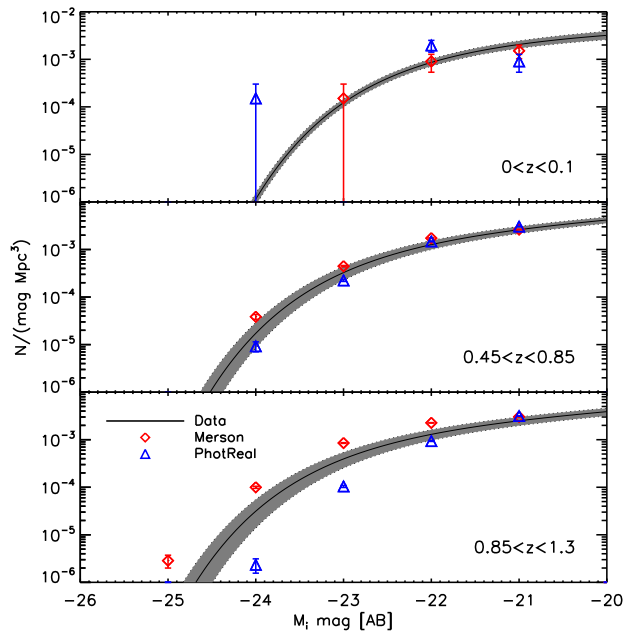} 
\caption{Rest-frame $i$ luminosity function for three different redshift bins: $0.<z<0.1$ (top panel), $0.45<z<0.85$ (model panel) and $0.85<z<1.3$ (bottom panel). The red diamonds refer to the magnitude counts of the initial mock catalogue and the blue triangles refer to the same counts for the \texttt{PhotReal} post-processed mock catalogue for the LSST survey. The solid line and the gray area are the observational luminosity function obtained by Montero-Dorta \& Prada (2009) for the low redshift bin and Gabasch et al. 2006 with the Case 3 evolution for the other higher redshift bins.}
\label{fig:LFlsst}
\end{figure}

We have also checked the concordance of the LF at lower redshift. In Fig. \ref{fig:LFlsst}, we show the $i$ LF for three different redshift bins as measured by   \cite{montero-dorta09} for the SDSS DR6 survey at low redshift (0$<z<$0.1)  and \cite{gabasch06} from the FORS Deep Field (FDF) survey for 0.45$<z<$1.3. As before, we overplot the luminosity counts of the initial and \texttt{PhotReal} post-processed LSST simulation for all the three redshift bins.  The counts obtained for both mock catalogues are comparable and reproduce well observational luminosity function at any redshift.

\subsection{Stellar mass function}
\label{sec:smasssec}

\begin{figure}
\centering
\includegraphics[clip,angle=0,width=1.0\hsize]{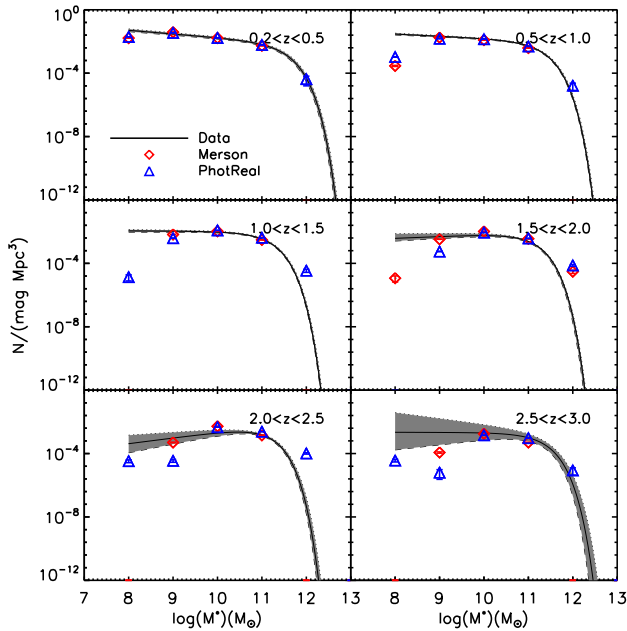} 
\caption{Stellar mass function for six different redshift bins: $0.2<z<0.5$ (top left panel), $0.5<z<1.0$ (top right panel), $1.0<z<1.5$ (middle left panel), $1.5<z<2.0$ (middle right panel), $2.0<z<2.5$ (bottom left panel) and $2.5<z<3.0$ (bottom right panel). The red diamonds refer to the stellar mass counts of the initial mock catalogue and the blue triangles refer to the same counts for the \texttt{PhotReal} post-processed mock catalogue for the Euclid-opt survey. The solid line and the gray area are the observational SM function obtained by Muzzin et al. 2013.}
\label{fig:SM}
\end{figure}

We have looked into the agreement of the observational stellar mass (SM) function with the one obtained from our simulations. In Fig. \ref{fig:SM}, we display the observational SM obtained by  \cite{muzzin13} from the COSMOS/UltraVista field within $0<z<3$ for six different redshift bins. As before, we overplotted the SM counts for the original mock photometry and the \texttt{PhotReal} post-processed photometry. We find a remarkable good agreement between the observational SM function and both estimations in the mock catalogues, giving also support to the fact that we do not substantially alter the initial mock catalogue.

\begin{figure}
\centering
\includegraphics[clip,angle=0,width=1.0\hsize]{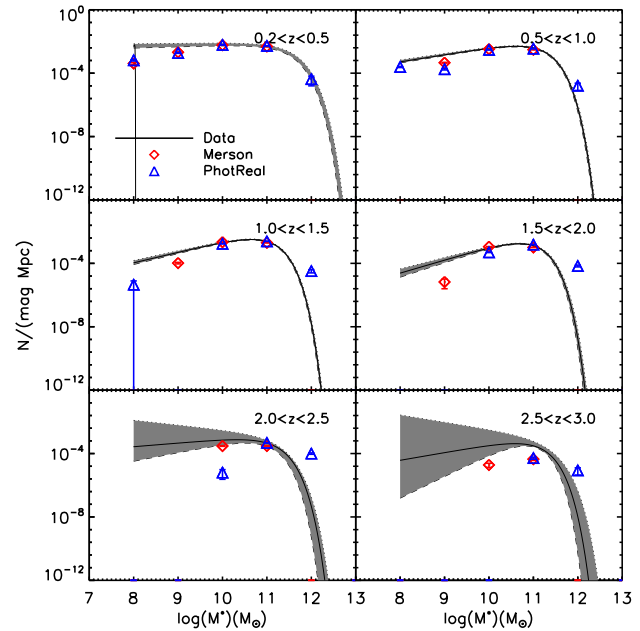} 
\caption{Stellar mass function for quiescent galaxies for six different redshift bins: $0.2<z<0.5$ (top left panel), $0.5<z<1.0$ (top right panel), $1.0<z<1.5$ (middle left panel), $1.5<z<2.0$ (middle right panel), $2.0<z<2.5$ (bottom left panel) and $2.5<z<3.0$ (bottom right panel). The red diamonds refer to stellar mass counts of those galaxies classified as early type ($t_b<3$) in the original mock catalogue fit. The blue triangles refer to the stellar mass counts of those galaxies classified as early type ($t_b<3$) in the \texttt{PhotReal} post-processed mock catalogue for the Euclid-opt survey. The solid line and the gray area are the observational quiescent SM function obtained by Muzzin et al. 2013.}
\label{fig:SMQ}
\end{figure}

\begin{figure}
\centering
\includegraphics[clip,angle=0,width=1.0\hsize]{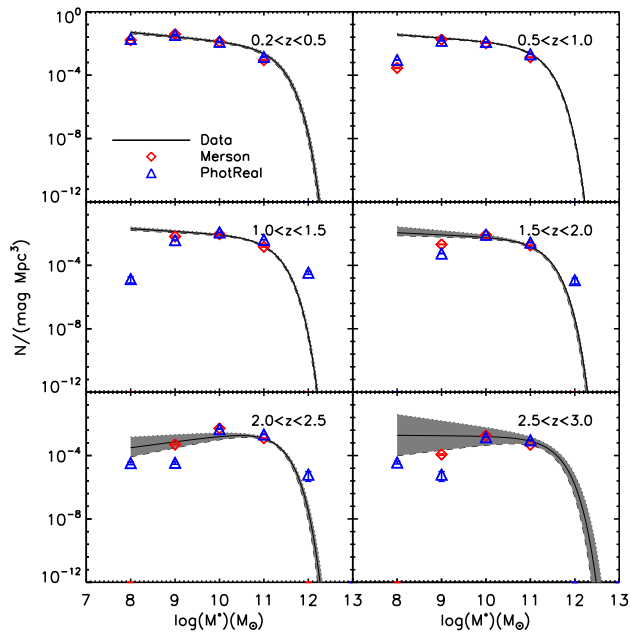} 
\caption{Stellar mass function for quiescent galaxies for six different redshift bins: $0.2<z<0.5$ (top left panel), $0.5<z<1.0$ (top right panel), $1.0<z<1.5$ (middle left panel), $1.5<z<2.0$ (middle right panel), $2.0<z<2.5$ (bottom left panel) and $2.5<z<3.0$ (bottom right panel). The red diamonds refer to stellar mass counts of those galaxies classified as late-type ($t_b>3$) in the original mock catalogue fit. The blue triangles refer to the stellar mass counts of those galaxies classified as late-type ($t_b>3$) in the \texttt{PhotReal} post-processed mock catalogue for the Euclid-opt survey. The solid line and the gray area are the observational quiescent SM function obtained by Muzzin et al. 2013.}
\label{fig:SMS}
\end{figure}

We have also compared the SM function for different galaxy types. In Figs. \ref{fig:SMQ} and \ref{fig:SMS}, we show observational SM functions by \cite{muzzin13} for quiescent galaxies and star forming galaxies respectively, we also display in these plots the SM counts for these types of galaxies for the original and \texttt{PhotReal} post-processed catalogue. To do this, we have defined quiescent galaxies  (star-forming galaxies) as those that the spectral classification obtained from the fit to the original $t_i$, or \texttt{PhotReal} post-processed photometry, $t_f$ has a spectral type  $<3$ ($\ge 3$). 

In general, we find a very good agreement between the observational SM function for the star-forming galaxies and the initial and  \texttt{PhotReal} post-processed SM distribution for any redshift bin. There exists also a good agreement between the observational SM function for the quiescent galaxies and both mock SM, at least for redshift higher than 1. Interestingly, the high mass end of the mass function obtained with  \texttt{PhotReal}  reproduces better the observational stellar mass function, compensating for the fact that the original mock catalogue shows lack of quiescent galaxies in this mass range.

\subsection{Angular correlation function}

As a final test for our mock catalogues, we have also compared the obtained angular correlation functions with observational data. For this purpose we used the public software \texttt{athena}\footnote{http://www.cosmostat.org/athena.html}. Since the basis of our cosmological simulation is the Millennium simulation and we have not perturbed the galaxy positions,  the overall correlation function should reflect reality. However, our aim is to test  if the angular correlation function of passive and star-forming galaxies has unrealistically changed when applying \texttt{PhotReal}.

\begin{figure}
\centering
\includegraphics[clip,angle=0,width=1.0\hsize]{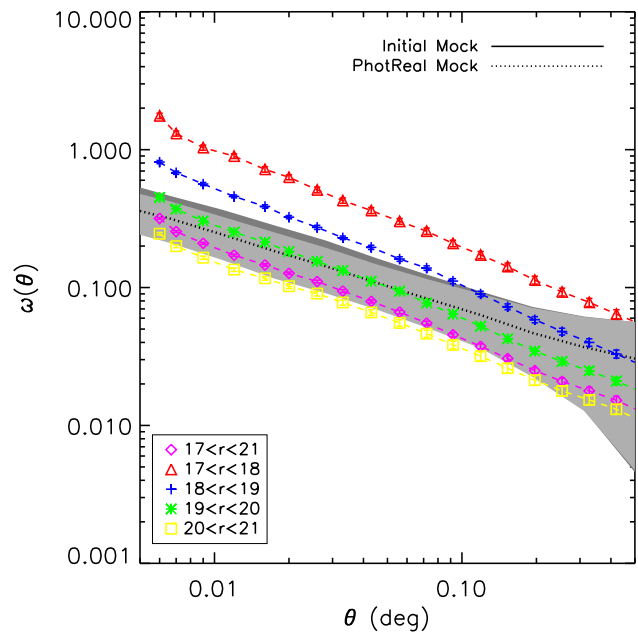} 
\caption{Angular correlation function for the original mock catalogue (solid line) and the \texttt{PhotReal} post-processed mock catalogue (dotted line) for the sample limited to redshift $z<0.2$. The shaded areas refer to 3$\sigma$ errors. The color symbols and the dashed lines show the observational SDSS angular correlation function calculated by Wang et al. 2013.}
\label{fig:CorrF}
\end{figure}

In Fig \ref{fig:CorrF}, we display the overall angular correlation function for the initial and the  \texttt{PhotReal} post-processed mock limited to redshift 0.2. This is compared to the angular correlation functions obtained from the SDSS \citep{wang13} split for different magnitude bins. For the sake of clarity, we have not split our sample in magnitude bins. However, we see that our estimation is well located within the different observational angular correlation functions.

\begin{figure}
\centering
\includegraphics[clip,angle=0,width=1.0\hsize]{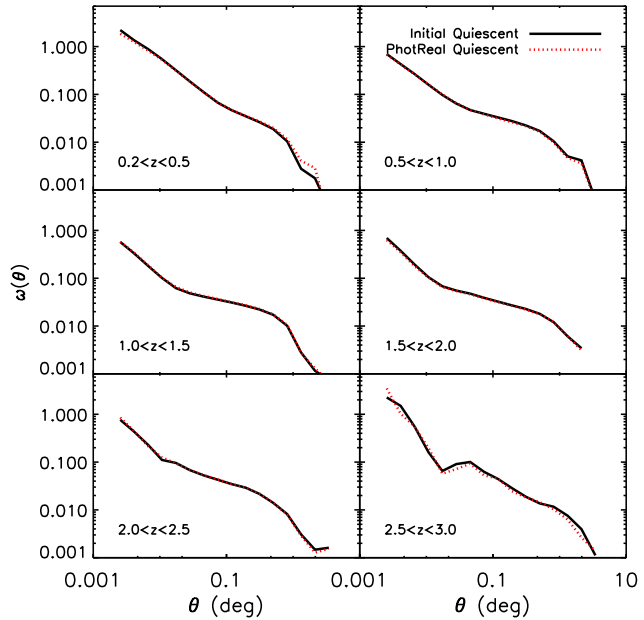} 
\caption{Angular correlation function for the original mock catalogue (solid line) and the \texttt{PhotReal} post-processed mock catalogue (red dotted line) for the quiescent galaxies for six different redshift bins: $0.2<z<0.5$ (top left panel), $0.5<z<1.0$ (top right panel), $1.0<z<1.5$ (middle left panel), $1.5<z<2.0$ (middle right panel), $2.0<z<2.5$ (bottom left panel) and $2.5<z<3.0$ (bottom right panel). See the text for a definition of quiescent galaxy for each sample.}
\label{fig:CorrFQ}
\end{figure}

\begin{figure}
\centering
\includegraphics[clip,angle=0,width=1.0\hsize]{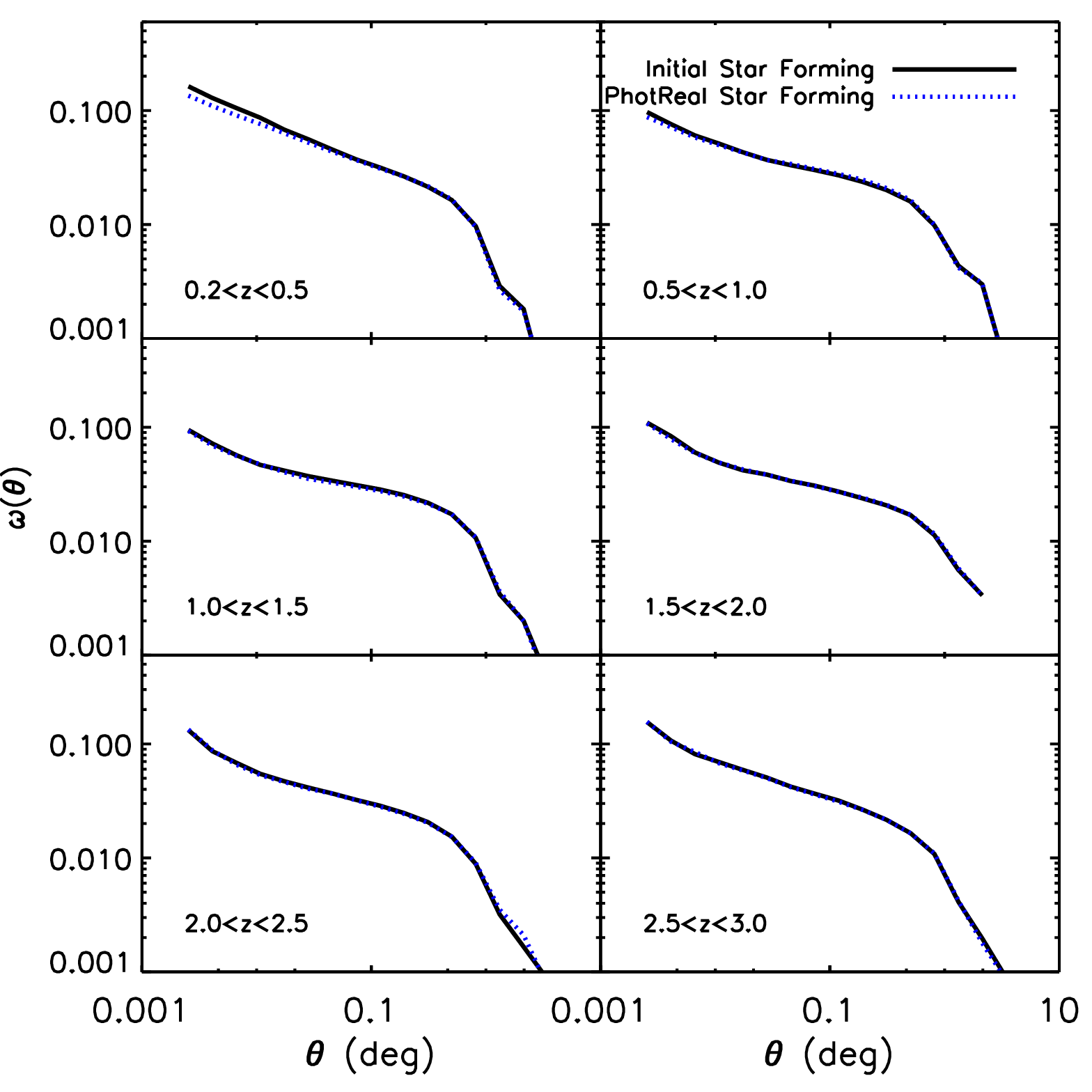} 
\caption{Angular correlation function for the original mock catalogue (solid line) and the \texttt{PhotReal} post-processed mock catalogue (blue dotted line) for the star-forming galaxies for six different redshift bins: $0.2<z<0.5$ (top left panel), $0.5<z<1.0$ (top right panel), $1.0<z<1.5$ (middle left panel), $1.5<z<2.0$ (middle right panel), $2.0<z<2.5$ (bottom left panel) and $2.5<z<3.0$ (bottom right panel). See the text for a definition of star-forming galaxy for each sample.}
\label{fig:CorrFS}
\end{figure}

Moreover, we have split the angular correlation functions for quiescent and star forming galaxies with the same criteria for classifying the initial and \texttt{PhotReal} post-processed photometry as in the previous section. In Figs. \ref{fig:CorrFQ} and \ref{fig:CorrFS}, we display the angular correlation function for the original and Euclid-Opt \texttt{PhotReal} mock catalogue for the  quiescent and star-forming galaxies respectively for six different redshift bins. We observe that the angular correlation functions for the both types of galaxies are almost indistinguible for the original and \texttt{PhotReal} post-processed catalogue, confirming the unalterability of the main properties of the mock galaxies.

\section{Photometric redshift performance}

In this section, we investigate the behavior of the photometric redshifts obtained from the mock catalogues described in section 3 for each next generation survey in terms of photometric redshift resolution and catastrophic outliers rate.

\subsection{Photometric Redshift Resolution}

In order to obtain a robust measurement of the mean photometric redshift resolution, we have used the NMAD estimator following a similar approach as in \cite{brammer08,molino14}. This estimator is known to be less sensitive to the behavior of the distribution. As such, the NMAD estimator is defined as:

\begin{equation}
\sigma_{NMAD}=1.48\times median\bigg(\frac{|\Delta z - median(\Delta z)|}{1+z_s}\bigg)
\end{equation}
where $\Delta z=z_b-z_s$

\begin{figure}
\centering
\includegraphics[clip,angle=0,width=1.0\hsize]{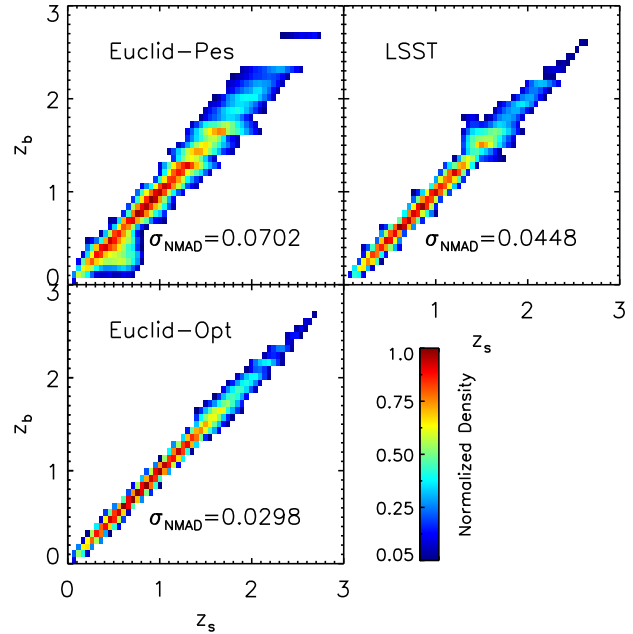} 
\caption{Density maps of spectroscopic redshift versus photometric redshift for each different survey considered for all the simulation (LSST, upper right; Euclid pessimistic, upper left and Euclid optimistic, bottom left). For a definition of Euclid pessimistic and optimistic case see \S 3.3. The color scale is logarithmic and redder colors represents higher densities while bluer colors refer to less dens regions. The overall photometrical redshift precision for each survey is quoted in each panel. Individually, LSST quotes the highest photometric redshift dispersion in the whole redshift range due to the drop of the flux in the optical bands at high redshift. The Euclid pessimistic and optimistic cases sample well all the redshift range up to redshift 3, having the later an improve in overall photometric redshift precision of a factor of two with respect to the former.}
\label{fig:zszb}
\end{figure}

In Fig \ref{fig:zszb}, we show the density maps of the photometric redshift versus spectroscopic redshift for the three cases considered. We immediately notice several differences between the considered surveys. In the case of the LSST, the lack of infrared data gets compensated with the depth of the optical data, being able to reach higher redshift and magnitude limits with a  photometric redshift precision of 0.045 for all the redshift range. We see some features at redshift $\sim$1.5, corresponding to the limit at which the spectrum of the objects is shifted to the infrared. The two cases considered for the Euclid surveys sample well the optical and infrared spectrum providing a less biased and lower rates of catastrophic outliers than the other surveys. The expected accuracy for the Euclid pessimistic and optimistic case is $\sigma_{NMAD}=0.070$ and $0.0298$ respectively. It is noticeable the higher dispersion found for the Euclid-Pes at lower redshift $<$0.4, which is probably due to the lack of UV data (i.e. \citealt{schmidt13}).

\begin{figure}
\centering
\includegraphics[clip,angle=0,width=1.0\hsize]{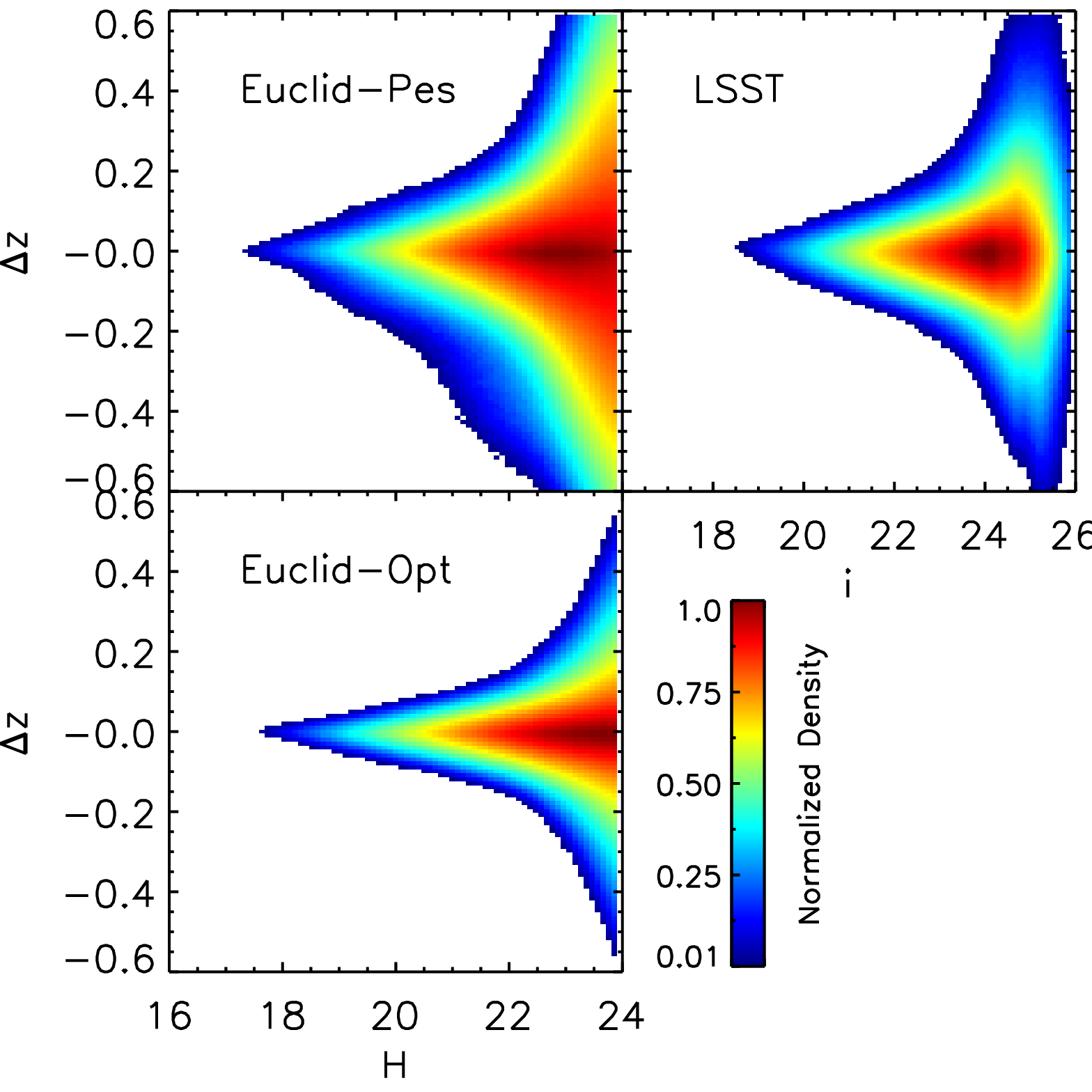} 
\caption{Density maps of magnitude versus the photometric bias  for the three different survey considered in this work (Euclid pessimistic, top left; Euclid optimistic, bottom left and LSST, top right).The color scale are the same as in Fig \ref{fig:zszb}. The axis scale are the same for all the panels. The behavior of the photometric redshifts for the LSST and both Euclid cases show larger dispersions and biases down to deeper magnitudes, having the Euclid Optimistic case a significantly smaller dispersion than the Pessimistic case and the LSST.}
\label{fig:mdeltaz}
\end{figure}

We have now investigated the photometric redshift performance as a function of the magnitude. The results are collected in Table \ref{tab:photozmag} and in Fig. \ref{fig:mdeltaz}, we show the behavior of the photometric redshift bias as a function of magnitude.  The parameter $\eta$ refers to the fraction of outliers and it is defined in \S\ref{sec:outliers}. Clearly, the number of bands covering a wide range of the spectrum has a direct impact in the photometric redshift accuracy and photometric redshift bias. While the Euclid-Opt has photometric redshift precision of $<0.02$ down to $H=24$ mag AB, we have an increase factor of 2-3 for the Euclid-Pes and the LSST cases for the same range of magnitudes. The same effect is observed in the photometric redshift bias.

\begin{table*}
      \caption{Photometric redshift comparison for different surveys as a a function of magnitude}
      \[
         \begin{array}{c|ccc|ccc|ccc}
		\hline
		\multicolumn{1}{c}{\rm }&
		\multicolumn{1}{c}{\rm }&
		\multicolumn{1}{c}{\rm LSST}&
		\multicolumn{1}{c}{} &
		\multicolumn{1}{c}{\rm }&
		\multicolumn{1}{c}{\rm Euclid-P}&
		\multicolumn{1}{c}{} &
		\multicolumn{1}{c}{\rm }&
		\multicolumn{1}{c}{\rm Euclid-O}&
		\multicolumn{1}{c}{} \\\hline\hline
		{\rm mag} &   {\rm \Delta z} & {\rm \sigma_{NMAD}} & {\rm \eta} &  {\rm \Delta z} &{\rm \sigma_{NMAD}} & {\rm \eta} &  {\rm \Delta z} & {\rm \sigma_{NMAD}} &{\rm \eta}   \\\hline
     16.0-16.5 &       0.0276 &     0.0333   &  15.3281 &       0.0239  &    0.0342  &   19.2475  &   0.0040  &    0.0137  &    0.2043    \\
     16.5-17.0   &     0.0242  &    0.0312   &  13.2013  &      0.0206  &    0.0324  &   15.9318  &   0.0041  &    0.0137  &    0.1542   \\
     17.0-17.5  &     0.0199   &   0.0290   &  10.9262   &      0.0157  &    0.0300  &   11.9770  &   0.0030  &    0.0143  &    0.1742   \\
     17.5-18.0   &    0.0176   &   0.0287   &   9.4708   &     0.0113   &   0.0309  &    9.8102    &  0.0033   &   0.0152   &   0.1491  \\
     18.0-18.5   &     0.0156   &   0.0277  &    7.7501   &   0.0063  &    0.0318    &  8.1134     &      0.0023    &  0.0152   &   0.1668   \\
     18.5-19.0   &     0.0124  &    0.0273  &    6.6554    &      0.0016   &   0.0322  &    7.2817     &      0.0014   &   0.0153 &     0.1645     \\
     19.0-19.5   &      0.0096  &    0.0268   &   5.5151  &       -0.0015    &  0.0315   &   7.1102   &    0.0008   &   0.0151  &    0.1601       \\
     19.5-20.0   &      0.0067 &     0.0260  &    4.6757  &       -0.0031   &   0.0300    &  7.2907    &      0.0007  &    0.0150  &    0.1496       \\
     20.0-20.5   &     0.0046   &   0.0251  &    3.8189   &        -0.0034   &   0.0287 &     7.4268   &      0.0009  &    0.0150  &    0.1319        \\     
     20.5-21.0   &      0.0029   &   0.0242   &   3.1238   &      -0.0037  &    0.0277  &    7.5001  &    0.0011    &  0.0150    &  0.1296        \\
     21.0-21.5   &      0.0015  &    0.0237  &    2.6430  &      -0.0040  &    0.0278 &     7.4401    &    0.0011   &   0.0150   &   0.1304          \\
     21.5-22.0   &    0.0006   &   0.0235   &   2.2978   &       -0.0046  &    0.0291  &    7.2443    &      0.0010   &   0.0154   &   0.1564         \\
     22.0-22.5   &     -0.0001   &    0.0236   &    2.1691   &     -0.0061   &    0.0323    &   7.1957      &      0.0006     &  0.0163   &    0.2341          \\
     22.5-23.0   &   -0.0005    &   0.0238   &    2.3066    &    -0.0089    &   0.0396   &    8.9712    &    -0.0001    &   0.0182   &    0.4693          \\
     23.0-23.5   &  -0.0008     &  0.0242    &   2.7025     &   -0.0123    &   0.0546   &   15.2487      &     -0.0006    &   0.0216     &  1.2641          \\     
     23.5-24.0   &     -0.0007 &     0.0252   &   3.5816   &     -0.0208   &   0.0791  &   26.1357  &    -0.0006    &  0.0277    &  3.5440 \\
     24.0-24.5   &   -0.0007  &    0.0278  &    5.2341     &      --      & --     &    --    &  --       & --    &     --\\
     24.5-25.0   &     -0.0005 &     0.0327   &   7.5035    &     --      & --     &    --    &  --       & --    &     --\\
     25.0-25.5   &   -0.0051  &    0.0441   &  12.9205     &   --      & --     &    --    &  --       & --    &     --\\
     25.5-26.0   &   0.0031   &   0.0657    & 23.2505      &    --      & --     &    --    &  --       & --    &     --\\\hline
 	\end{array}
      \]
\label{tab:photozmag}
   \end{table*}
     
\begin{figure}
\centering
\includegraphics[clip,angle=0,width=1.0\hsize]{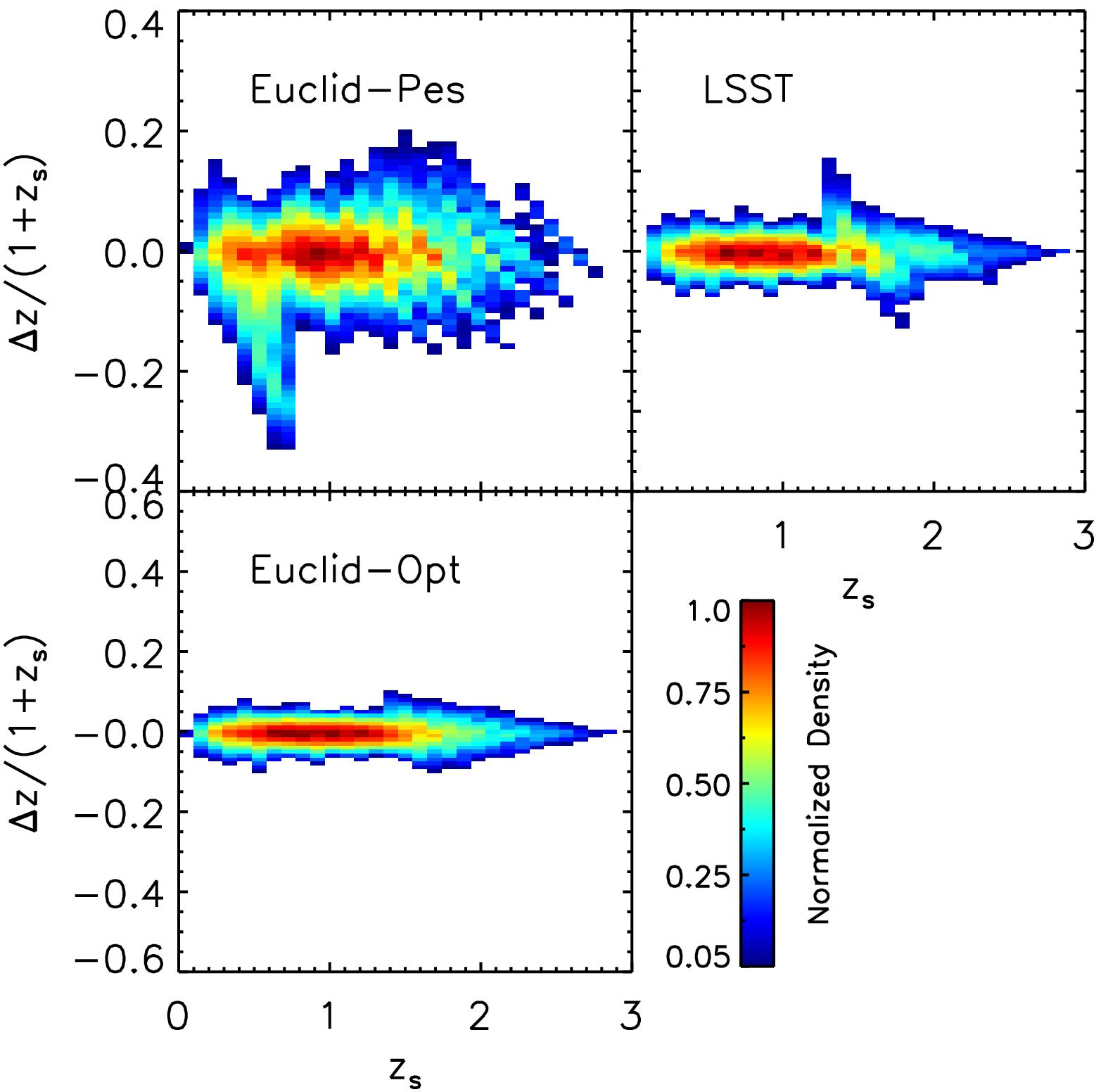} 
\caption{Density maps of redshift versus the photometric bias for the LSST, upper right; Euclid pessimistic, upper left and Euclid optimistic, bottom left. The color scale are the same as in Fig \ref{fig:zszb}. The axis scale are the same for all the panels.  The behavior for the LSST and Euclid-Pessimistic case becomes similar (less than a factor of 2) up to redshift 1.5. At higher redshift, the bias for the Euclid-Pessimistic case. than the Euclid Optimistic case, which obtains an overall good performance ($\sigma_{NMAD}\sim 0.02-0.4$) and negligible photometric redshift bias up to redshift 3.}
\label{fig:zdeltaz}
\end{figure}

We have also looked into the performance of the photometric redshifts for each different survey as a function of the redshift. The results are summarized in Table \ref{tab:photozz} and are displayed in Fig.  \ref{fig:zdeltaz}. We find that the photometric redshift dispersion is always at least a factor of 1.5 higher for the Euclid-Pessimistic than for the LSST. This result remarks that the presence of the infrared bands do not improve the photometric redshift resolution at $z<1$. By contrast,  the improvement in the photometric redshift performance found in using a deeper optical counterpart becomes remarkable when comparing the results of the two Euclid cases. The photometric redshift resolution of the Euclid-Optimistic is a factor of 2 to 5 better than in the Pessimistic case both in terms of photometric redshift accuracy and bias. In general, we find that the photometric redshift dispersion depends more strongly on magnitude than on redshift, in agreement with other works \citep{hildebrandt12,dahlen13,raichoor14,molino14}.

\begin{table*}
      \caption{Photometric redshift comparison for different surveys as a a function of redshift}
      \[
         \begin{array}{c|ccc|ccc|ccc}
		\hline
		\multicolumn{1}{c}{\rm }&
		\multicolumn{1}{c}{\rm }&
		\multicolumn{1}{c}{\rm LSST}&
		\multicolumn{1}{c}{} &
		\multicolumn{1}{c}{\rm }&
		\multicolumn{1}{c}{\rm Euclid-P}&
		\multicolumn{1}{c}{} &
		\multicolumn{1}{c}{\rm }&
		\multicolumn{1}{c}{\rm Euclid-O}&
		\multicolumn{1}{c}{} \\\hline\hline
		{\rm z} &   {\rm \Delta z} & {\rm \sigma_{NMAD}} & {\rm \eta} &  {\rm \Delta z} &{\rm \sigma_{NMAD}} & {\rm \eta} &  {\rm \Delta z} & {\rm \sigma_{NMAD}} &{\rm \eta}   \\\hline
      0.0-0.5  &       0.0057 &     0.0302   &   5.8276  &    0.0011   &   0.0457 &    13.6601  &    0.0009   &   0.0214    &  1.4644\\
      0.5-1.0  &   -0.0021  &    0.0237   &   2.0035   &  -0.0225   &   0.0449    & 17.8672 &    -0.0007    &  0.0174  &    1.0415\\
     1.0-1.5  &     0.0143   &   0.0300   &  10.3601  &    0.0023  &    0.0414   &  12.4176 &     0.0013    &  0.0171   &   0.8455\\
      1.5-2.0 &     -0.0199   &   0.0452  &   11.0907   &   0.0089   &   0.0564   &  15.1593  &   -0.0002   &   0.0291  &    2.4469\\
      2.0-2.5  &        -0.0409   &   0.0401 &    12.4108    & -0.0429   &   0.0581   &  14.7191   &  -0.0017 &     0.0285  &    3.0074\\
      2.5-3.0  &    -0.0181   &   0.0229  &    7.2210  &   -0.1783  &    0.0577 &    20.8091   &  -0.0048   &   0.0166    &  1.8212 \\\hline
	\end{array}
      \]
\label{tab:photozz}
   \end{table*}
   
\subsection{Fraction of Outliers}
\label{sec:outliers}

The term 'outliers' refers to those galaxies that, due to the well-known color-redshift degeneracies, gets assigned a very similar color to a theoretical galaxy at a very different redshift. Even if the Bayesian probability alleviates this problem due to the introduction of a Bayesian prior \citep{benitez00,bellagamba12}, the existence of such galaxies still remains and we need to quantify it.

One necessary requirement for good quality photometric redshifts is to have a small percentage of the total sample of galaxies becoming 'outliers'. We define fraction of outliers as in \cite{hildebrandt10}:

\begin{equation}
\eta=\frac{|\Delta z|}{1+z_s}>0.15
\end{equation}

\begin{figure}
\centering
\includegraphics[clip,angle=0,width=1.0\hsize]{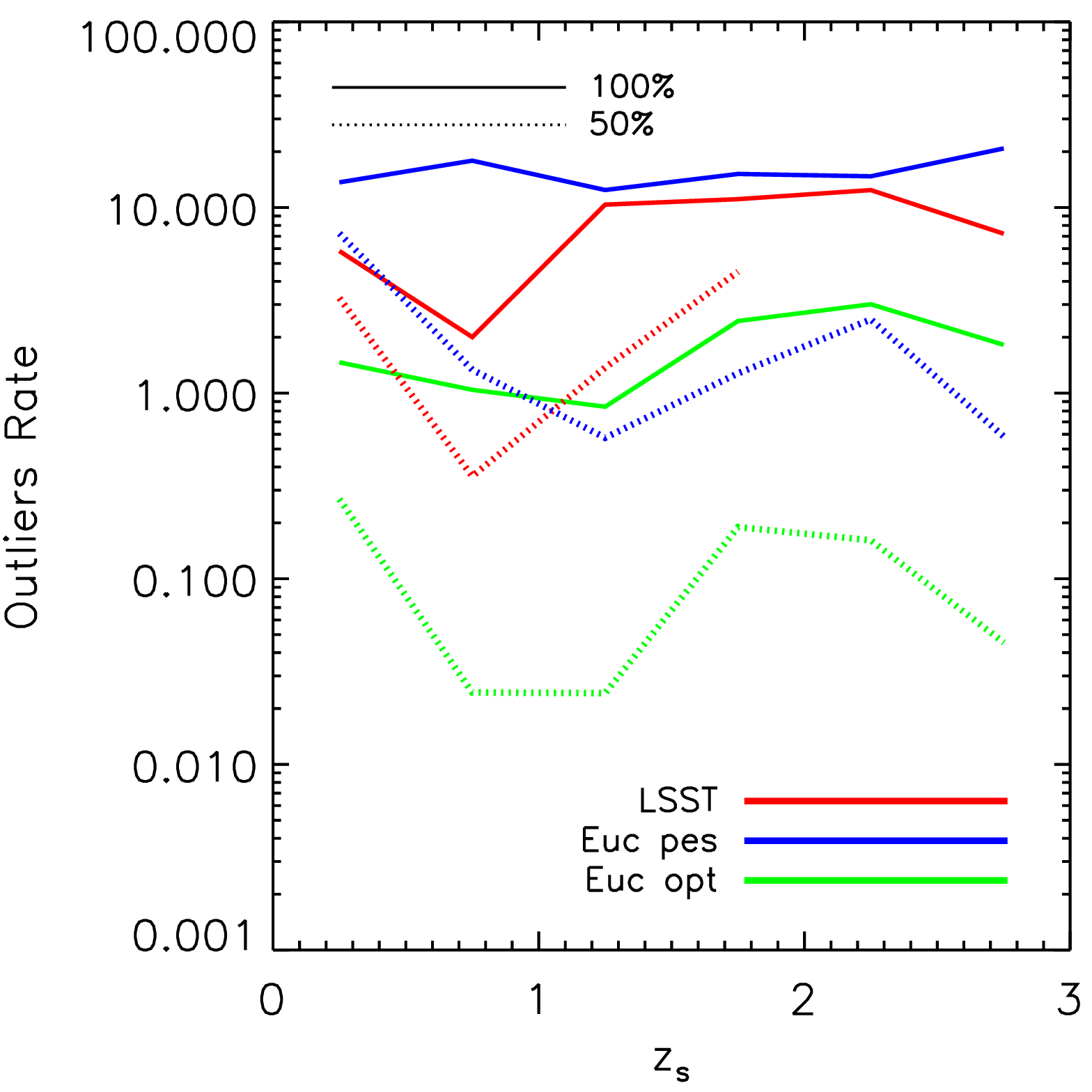} 
\caption{Fraction of outliers as a function of the redshift for the three different survey considered in this work (LSST, red line; Euclid pessimistic, blue line and Euclid optimistic, green line). The solid lines refer to the overall population for each survey while the dotted lines refer to the 50\% of the population with highest \emph{odds} for each different survey. In general, the fraction of outliers increases gently with redshift for all the surveys considered here, obtaining the lower rates for Euclid Optimistic, Euclid Pessimistic, and LSST in increasing order. If we consider the best 50\% quality \emph{odds} sample, the outliers rate decreases more than a factor of 10.}
\label{fig:zcatout}
\end{figure}

The results obtained for different surveys as a function of magnitude and redshift are collected in the third column of each survey in Tables \ref{tab:photozmag} and  \ref{tab:photozz} respectively. In Fig  \ref{fig:zcatout}, we show the outlier rate as a function of redshift for the different surveys.

While the Euclid-Opt survey has a very lower rate of catastrophic outliers as a function of magnitude, being less than 1\% down to $H=23$ mag, the LSST and Euclid-Pes shows a higher rate of outliers between 3-12\%. This separation illustrates that the multiple ($>12$) band surveys are much more effective in decreasing the rate of catastrophic outliers than surveys with fewer bands.

Looking at the same rate as a function of redshift, we see a milder dependence of the outlier rate and redshift. The rate of outliers remains close to $\sim$1-2\% for the Euc-Opt, $\sim$ 2-10\% for the LSST and$ \sim$ 10-20\% for the Euc-Pes survey. As before, the presence of the deep infrared data and the combination with deep optical data is a strong combination to decrease the rate of catastrophic outliers.

\section{Photometric redshift quality cuts}

As we have illustrated in the previous section, the existence of outliers in a photometric survey is unavoidable but can be quantified and keep under a moderate threshold. In addition, we can use one main advantage of some photometric redshift packages such as \texttt{BPZ2.0}, which is the fact that all the galaxies have assigned a redshift probability function. The study of this function provides a measure of the quality of the photometric redshift: very well behaved photometry redshifts will probably have a well defined narrow peak around the spectroscopic redshift of the galaxy, whereas galaxies with degeneracies will be assigned a multi peak probability distribution. 

\cite{benitez00} introduced the \emph{odds} parameter: a measurement of the quality of a photometric redshift. This \emph{odds} parameter is obtained by integrating the redshift probability function of the galaxy within 2$\sigma$ of its more probable value. These parameters has been widely used to select the best quality photometric redshift sample to study a number of different features (e.g. \citealt{coe06,benitez09a,benitez09b,bellagamba12,molino14,ascaso15}) and allows to improve the overall photometric redshift precision of the sample by previously modeling the selection bias.

In this section, we discuss the benefit obtained in terms of the improvement of the photometric redshift accuracy and the decrement of the outlier rates when using this parameter to select the best quality photometric redshift sample for a particular survey.
 
\subsection{Improvement in the photometric redshift resolution}

\begin{figure}
\centering
\includegraphics[clip,angle=0,width=1.0\hsize]{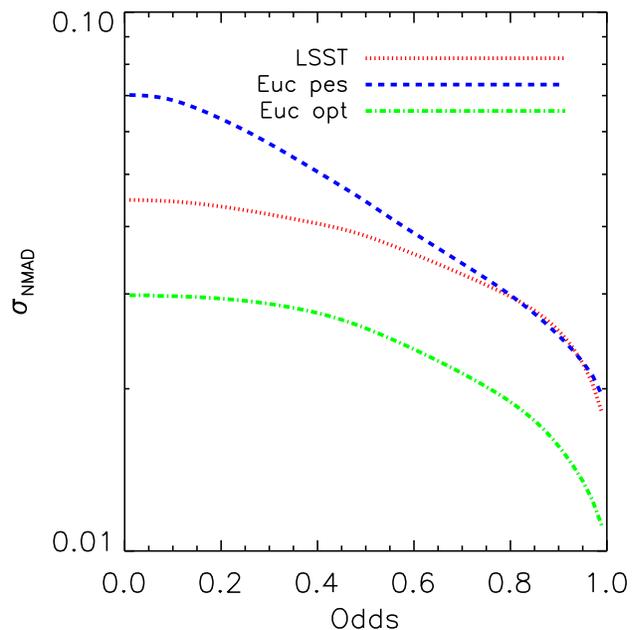} 
\caption{\emph{Odds} photometric redshift quality parameter versus photometric redshift dispersion for the three different survey considered in this work (LSST, red dotted line; Euclid pessimistic, blue dashed line and Euclid optimistic, green dotted-dashed line). Performing a safe quality \emph{odds} parameter in each of these surveys can decrease the photometric redshift dispersion significantly.}
\label{fig:oddsigmaz}
\end{figure}

In order to illustrate the increase in photometric redshift resolution by effectuating a cut in the \emph{odds} parameter, we plot in Fig. \ref{fig:oddsigmaz}, the photometric redshift accuracy as a function of the \emph{odds} parameter for each of the three different surveys considered in this study. As expected, we see a decrease of the photo-z dispersion as the \emph{odds} increase for all three surveys. Indeed, we can select a subsample of the accuracy desired by performing an accurate \emph{odds} cut. For instance, while the overall accuracy of LSST is 0.045, we could obtain a subsample of twice this accuracy (i.e. $\sigma_{NMAD} \sim 0.0279$) by selecting the 25\% highest \emph{odds} subsample. 

\begin{figure}
\centering
\includegraphics[clip,angle=0,width=1.0\hsize]{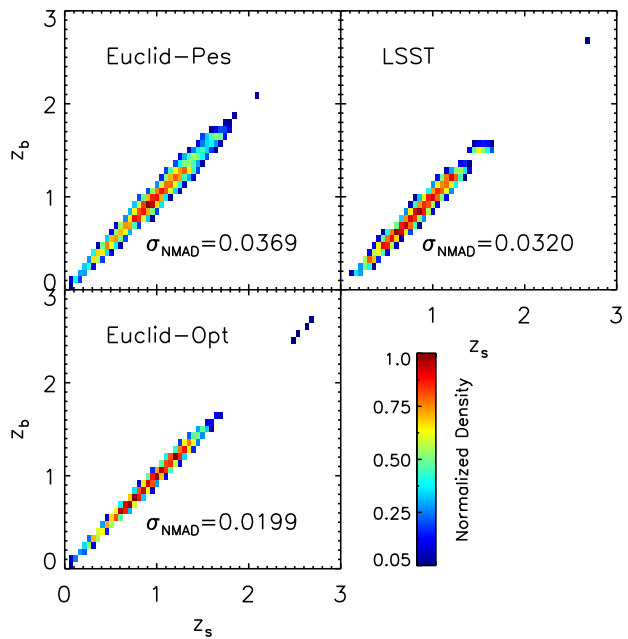} 
\caption{Density maps of spectroscopic redshift versus photometric redshift for the 50\% best quality photometric redshift for each different survey considered for all the simulation (LSST, upper right; Euclid pessimistic, bottom left and Euclid optimistic, bottom right).  The color scale are the same as in Fig \ref{fig:zszb}.  The overall photometrical redshift precision for each survey is quoted in each panel. The photometric redshift resolution increases between a factor of 1.5 and 3 for all surveys by selecting a best quality photometric redshift subsample, being particularly affected the density regions of outliers.}
\label{fig:zszbodd}
\end{figure}

We also illustrate the effect of this selection by plotting the analogous plot to the Fig. \ref{fig:zszb} in Fig. \ref{fig:zszbodd}, the behavior of the photometric redshift as a function of spectroscopic redshift for the 50\% best quality photometric redshifts for each of the surveys considered. The number quoted in the legend of each of the panels refers to the overall precision achieved, being this a factor between 1.5 and 3 with respect to the overall sample. In particular, a large fraction of outliers, departing from the one-to-one line in Fig. \ref{fig:zszb}, disappears by considering only galaxies with high odds.

\subsection{Decrease of outlier rates}

As mentioned in the previous section, in Fig. \ref{fig:zcatout}, we have illustrated the relation between the outlier rates as a function of redshift with solid lines for each of the different surveys. In this Figure, we have also plotted the outlier rates resulting for the 50\% best photometric redshift quality subsample for every survey. We plainly observe a dramatic descend of the outlier rate for each survey. The new rates are decreased by a factor of 3 to 10, putting in evidence the virtues of this selection.

When working in a particular science topic, one needs to understand the nature of any possible selection based on the  \emph{odds} parameter and quantify the possible bias introduced on the final scientific results. Once this effect is taken into account, the improvement in photometric precision and the decrease of outliers rate can be fully exploited and we encourage researchers to exploit this feature.

\section{Discussion and conclusions}

The advent of several large deep surveys is imminent and the need of precise tools to describe accurately their properties is pressing. In this work, we have developed mock catalogues mimicking realistically two of the up-coming generation stage IV surveys: LSST and Euclid (considering two cases for the optical counterpart, optimistic and pessimistic). We have confirmed that the main properties of this new mock reflect reality in terms of photometry, photometric errors, stellar masses, spectral types and photometric redshifts by comparing with observable relations. We have also provided predictions in terms of photometric redshift performance and selection cuts from the same basis for each survey.  The mocks will become publicly available\footnote{http://photmocks.obspm.fr/} and they will become extremely useful for making internal comparisons between the different surveys.  In the appendix, we include a description of the content of the mock catalogues.

Each of these surveys will provide output data observed with very different strategies. In this work, we have disentangled some of the advantages of using each of these different datasets in terms of photometric redshift resolution. First, the number of bands used to sample the optical and/or infrared spectrum  is directly related to the level of photometric redshift accuracy and photometric redshift bias as confirmed by other works (e.g. \citealt{benitez09a} and references herein). On the other hand, deep broad-band optical surveys allow us to sample the luminosity function in the optical range even if with worse photometric redshift resolution. They are also an excellent complement for deep infrared surveys, particularly to dramatically decrease the rate of outliers and improve the photometric redshift accuracy. Finally, the deep infrared surveys provide an excellent photometric redshift accuracy in a wide range of redshift ($z<3$) with a moderate rate of catastrophic outliers. Thus, the future combination of all these strategies can lead to large amount of ancillary data which will provide excellent resolution for many scientific purposes. These mocks will become extremely useful for making internal comparisons between different surveys.

\begin{figure}
\centering
\includegraphics[clip,angle=0,width=1.0\hsize]{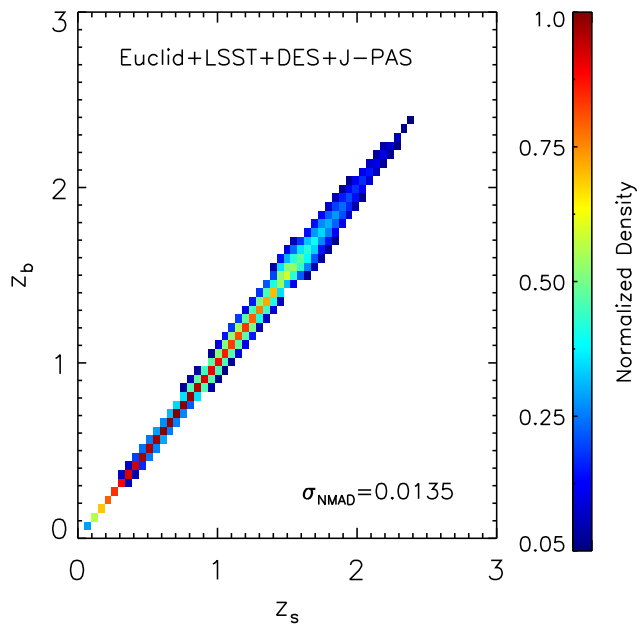} 
\caption{Density map of spectroscopic redshift versus photometric redshift  for a future 'super survey' being Euclid+LSST+DES+J-PAS. The color scale are the same as in Fig \ref{fig:zszb}.  The behavior of the photometric redshift when the ancillary data of some of the most important next generation survey is available would be excellent. The catastrophic outlier rate will decrease dramatically and the $0<z<3$ redshift range will be measured with an accuracy of 0.0095 without imposing any \emph{odds} quality cut.}
\label{fig:zszbALL}
\end{figure}

As stated in the introduction, a separate work (Ascaso et al. in prep) will be fully devoted to explore the photometric redshift expectations for J-PAS, a multiple narrow-band survey. We have then examined the performance of the photometric redshifts for a new 'super-survey' consisting in combining all the expected data for these surveys once they are made public and we are allowed to combine it. This 'super survey' will consist then in the combination of Euclid+LSST+DES+J-PAS and the results should account for all the advantages described through this work. In Fig \ref{fig:zszbALL}, we plot the density map of the photometric redshift versus the spectroscopic redshift for this 'super survey'.  Let's note that these three surveys will not overlap in all parts of the sky and their overlap will also be different in the southern and northern hemispheres.

\begin{figure}
\centering
\includegraphics[clip,angle=90,width=1.0\hsize]{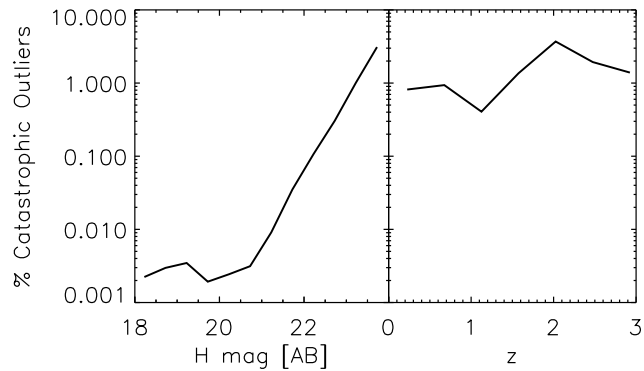} 
\caption{Catastrophic outliers rate as function of H-band (left plot) and redshift (right plot) for a future 'super survey' being Euclid+LSST+DES+J-PAS.  The catastrophic outlier rate will never be higher than  3\% in all magnitude and redshift ranges.}
\label{fig:outALL}
\end{figure}

The expected overall photometric redshift resolution for all the redshift range within $0<z<3$ and $i<27.5$ AB  will be of $\sigma_{NMAD}=0.0135$, with almost negligible photometric redshift bias. As for the rate of catastrophic outliers, we plot in Fig. \ref{fig:outALL}, the expected outlier rate for this 'super-survey' as a function of magnitude and redshift. As we see, this survey will keep the outliers rate below 1\%  up to redshift $\sim$2 and below 3\% for all the redshift and magnitude range. 

Hence, this future 'super survey' will support almost all the scientific cases for the community, becoming a superior machine comparable to a full spectroscopic survey ranging a very large range of redshift and magnitude.

In a forthcoming paper (Ascaso et al., in prep), we will continuos this work within the frame of \emph{Apples to Apples, A$^2$} to explore and compare the expected selection function for cluster and group detections for the three surveys considered here. We will also publish a third paper providing cosmological constraints from these cluster counts for different surveys. These series of works will help setting light on the expected predictions of these surveys and the mocks will be useful to set requirements to the science derived from them.

\section*{Acknowledgments}
We acknowledge the anonymous referee who provided useful suggestions to our paper. BA acknowledges financial support for a postdoctoral fellowship from the Observatory of Paris. SM acknowledge financial support from the Institut Universitaire de France (IUF), of which she is senior member. We acknowledge Alex Merson and Carlton Baugh for useful discussions and comments to our work. We also thank Michael Brown and Stefano Andreon for suggestions. BA acknowledges Rossella Licitra and Anand Raichoor for multiple discussions that helped to improve this paper.

\appendix
\section{Mock catalogues description.}
\label{release}
In this appendix we include the description of the content of the content of the galaxy mock catalogues in detail.  The three mock catalogues are 500 $deg^2$ wide and the magnitudes limits for the different bands are:

\begin{itemize}
\item Euclid: magnitude limit of H=24, J=24 and Y=24, down to 5$\sigma$, \citep{laureijs11}. 
\item LSST: magnitude limit of u=26.1, g=27.4, r=27.5, i=26.8, z=26.1 and y=24.9, down to 5$\sigma$, \citep{ivezic08}. 
\item DES: magnitude limit of g=25.2, r=24.8, i=24.0, z=23.4 and y=21.7, down to 10$\sigma$, \citep{mohr12}. 
\end{itemize}

\begin{table*}
\caption{Content of the Euclid Optimistic Mock Catalogue}
\begin{center}
\begin{tabular}{|l|c|c|c|c|c|c|c|}
\hline
\hline
& PARAMETER  &  DESCRIPTION \\
\hline
&      DHaloID$^*$	                  &  ID of the dark matter halo in the Millennium Simulation	 \\
&      GalaxyID$^*$	                  &  ID of the galaxy  \\
&      is\_central$^*$                  &  =1 if the galaxy is central, 0= satellite	 \\
&      MHalo$^*$	                  &  Mass of the host halo ($M_{\odot}$) 	 \\
&      MStars\_tot$^*$	                  &  Total Stellar mass of the galaxy ($M_{\odot}$) 	 \\
&      RA$^*$	                                    &  Right Ascension in decimal degrees [J2000]   \\
&      DEC$^*$	                           &  Declination in decimal degrees [J2000]   \\
&      zs$^*$                        &   Spectroscopic redshift \\
&       SDSS\_u\_or$^*$  	          &  SDSS\_u original magnitude  \\
&       SDSS\_g\_or$^*$ 	          &  SDSS\_g original magnitude  \\
&       SDSS\_r\_or$^*$	          &  SDSS\_r original magnitude  \\
&       SDSS\_i\_or$^*$ 	          &  SDSS\_i original magnitude  \\
&       SDSS\_z\_or$^*$ 	          &  SDSS\_z original magnitude \\
&       DES\_g\_or$^*$ 	          &  DES\_g original magnitude  \\
&       DES\_r\_or$^*$ 	          &  DES\_r original magnitude  \\
&       DES\_i\_or$^*$	          &  DES\_i original magnitude  \\
&       DES\_z\_or$^*$ 	          &  DES\_z original magnitude  \\
&       DES\_y\_or$^*$ 	          &  DES\_y original magnitude  \\
&       Euclid\_H\_or$^*$ 	          &  Euclid\_H original magnitude  \\
&       Euclid\_J\_or$^*$ 	          &  Euclid\_J original magnitude  \\
&       Euclid\_Y\_or$^*$ 	          &  Euclid\_Y original magnitude \\
\hline
&     zb                     & BPZ2.0 most likely redshift  \\
&     zb\_min             &  Lower limit (95p confidence) \\
&     zb\_max             & Upper limit (95p confidence) \\
&     tb                      &  BPZ2.0 most likely spectral type \\
&     Odds                & P(z) integrated within zb - 2$\sigma_{NMAD}$ and zb + 2$\sigma_{NMAD}$\\
&     Chi2                   &  Poorness of BPZ fit: observed vs. model fluxes \\ 
&     Stell\_Mass     &     \texttt{PhotReal}Stellar Mass (log10($M_{\odot}$))  \\
&       DES\_g, e\_DES\_g 	          &  DES\_g \texttt{PhotReal} magnitude \& uncertainty \\
&       DES\_r, e\_DES\_r 	          &  DES\_r \texttt{PhotReal} magnitude \& uncertainty \\
&       DES\_i, e\_DES\_i 	          &  DES\_i  \texttt{PhotReal} magnitude \& uncertainty \\
&       DES\_z, e\_DES\_z 	          &  DES\_z  \texttt{PhotReal} magnitude \& uncertainty\\
&       DES\_y, e\_DES\_y 	          &  DES\_y  \texttt{PhotReal} magnitude \& uncertainty \\
&       LSST\_u, e\_LSST\_u 	          &  LSST\_u  \texttt{PhotReal} magnitude \& uncertainty \\
&       LSST\_g, e\_LSST\_g 	          &  LSST\_g  \texttt{PhotReal} magnitude \& uncertainty \\
&       LSST\_r, e\_LSST\_r	          &  LSST\_r  \texttt{PhotReal} magnitude \& uncertainty \\
&       LSST\_i, e\_LSST\_i 	          &  LSST\_i  \texttt{PhotReal} magnitude \& uncertainty \\
&       LSST\_z, e\_LSST\_z 	          &  LSST\_z  \texttt{PhotReal} magnitude \& uncertainty \\
&       LSST\_y, e\_LSST\_y 	          &  LSST\_y  \texttt{PhotReal} magnitude \& uncertainty \\
&       Euclid\_H, e\_Euclid\_H 	          &  Euclid\_H  \texttt{PhotReal} magnitude \& uncertainty \\
&       Euclid\_J, e\_Euclid\_J 	          &  Euclid\_J  \texttt{PhotReal} magnitude \& uncertainty \\
&       Euclid\_Y, e\_Euclid\_Y 	          &  Euclid\_Y  \texttt{PhotReal} magnitude \& uncertainty \\
\hline
\hline
\end{tabular}
\end{center}
$^*$ as in the original Merson et al. mock catalogue
\end{table*}

\begin{table*}
\caption{Content of the Euclid Pessimistic Mock Catalogue}
\begin{center}
\begin{tabular}{|l|c|c|c|c|c|c|c|}
\hline
\hline
& PARAMETER  &  DESCRIPTION \\
\hline
&      DHaloID$^*$	                  &  ID of the dark matter halo in the Millennium Simulation	 \\
&      GalaxyID$^*$	                  &  ID of the galaxy  \\
&      is\_central$^*$                  &  =1 if the galaxy is central, 0= satellite	 \\
&      MHalo$^*$	                  &  Mass of the host halo ($M_{\odot}$) 	 \\
&      MStars\_tot$^*$	                  &  Total Stellar mass of the galaxy ($M_{\odot}$) 	 \\
&      RA$^*$	                                    &  Right Ascension in decimal degrees [J2000]   \\
&      DEC$^*$	                           &  Declination in decimal degrees [J2000]   \\
&      zs$^*$                        &   Spectroscopic redshift \\
&       SDSS\_u\_or$^*$  	          &  SDSS\_u original magnitude  \\
&       SDSS\_g\_or$^*$ 	          &  SDSS\_g original magnitude  \\
&       SDSS\_r\_or$^*$	          &  SDSS\_r original magnitude  \\
&       SDSS\_i\_or$^*$ 	          &  SDSS\_i original magnitude  \\
&       SDSS\_z\_or$^*$ 	          &  SDSS\_z original magnitude \\
&       DES\_g\_or$^*$ 	          &  DES\_g original magnitude  \\
&       DES\_r\_or$^*$ 	          &  DES\_r original magnitude  \\
&       DES\_i\_or$^*$	          &  DES\_i original magnitude  \\
&       DES\_z\_or$^*$ 	          &  DES\_z original magnitude  \\
&       DES\_y\_or$^*$ 	          &  DES\_y original magnitude  \\
&       Euclid\_H\_or$^*$ 	          &  Euclid\_H original magnitude  \\
&       Euclid\_J\_or$^*$ 	          &  Euclid\_J original magnitude  \\
&       Euclid\_Y\_or$^*$ 	          &  Euclid\_Y original magnitude \\
\hline
&     zb                     & BPZ2.0 most likely redshift  \\
&     zb\_min             &  Lower limit (95p confidence) \\
&     zb\_max             & Upper limit (95p confidence) \\
&     tb                      &  BPZ2.0 most likely spectral type \\
&     Odds                & P(z) integrated within zb - 2$\sigma_{NMAD}$ and zb + 2$\sigma_{NMAD}$\\
&     Chi2                   &  Poorness of BPZ fit: observed vs. model fluxes \\ 
&     Stell\_Mass     &     \texttt{PhotReal} Stellar Mass (log10($M_{\odot}$))  \\
&       DES\_g, e\_DES\_g 	          &  DES\_g  \texttt{PhotReal} magnitude \& uncertainty \\
&       DES\_r, e\_DES\_r 	          &  DES\_r  \texttt{PhotReal} magnitude \& uncertainty \\
&       DES\_i, e\_DES\_i 	          &  DES\_i  \texttt{PhotReal} magnitude \& uncertainty \\
&       DES\_z, e\_DES\_z 	          &  DES\_z  \texttt{PhotReal} magnitude \& uncertainty \\
&       DES\_y, e\_DES\_y 	          &  DES\_y  \texttt{PhotReal} magnitude \& uncertainty \\
&       Euclid\_H, e\_Euclid\_H 	          &  Euclid\_H  \texttt{PhotReal} magnitude \& uncertainty \\
&       Euclid\_J, e\_Euclid\_J 	          &  Euclid\_J  \texttt{PhotReal} magnitude \& uncertainty \\
&       Euclid\_Y, e\_Euclid\_Y 	          &  Euclid\_Y  \texttt{PhotReal} magnitude \& uncertainty \\
\hline
\hline
\end{tabular}
\end{center}
$^*$ as in the original Merson et al. mock catalogue
\end{table*}

\begin{table*}
\caption{Content of the LSST Mock Catalogue}
\begin{center}
\begin{tabular}{|l|c|c|c|c|c|c|c|}
\hline
\hline
& PARAMETER  &  DESCRIPTION \\
\hline
&      DHaloID$^*$	                  &  ID of the dark matter halo in the Millennium Simulation	 \\
&      GalaxyID$^*$	                  &  ID of the galaxy  \\
&      is\_central$^*$                  &  =1 if the galaxy is central, 0= satellite	 \\
&      MHalo$^*$	                  &  Mass of the host halo ($M_{\odot}$) 	 \\
&      MStars\_tot$^*$	                  &  Total Stellar mass of the galaxy ($M_{\odot}$) 	 \\
&      RA$^*$	                                    &  Right Ascension in decimal degrees [J2000]   \\
&      DEC$^*$	                           &  Declination in decimal degrees [J2000]   \\
&      zs$^*$                        &   Spectroscopic redshift \\
&       SDSS\_u\_or$^*$  	          &  SDSS\_u original magnitude  \\
&       SDSS\_g\_or$^*$ 	          &  SDSS\_g original magnitude  \\
&       SDSS\_r\_or$^*$	          &  SDSS\_r original magnitude  \\
&       SDSS\_i\_or$^*$ 	          &  SDSS\_i original magnitude  \\
&       SDSS\_z\_or$^*$ 	          &  SDSS\_z original magnitude \\
&       DES\_g\_or$^*$ 	          &  DES\_g original magnitude  \\
&       DES\_r\_or$^*$ 	          &  DES\_r original magnitude  \\
&       DES\_i\_or$^*$	          &  DES\_i original magnitude  \\
&       DES\_z\_or$^*$ 	          &  DES\_z original magnitude  \\
&       DES\_y\_or$^*$ 	          &  DES\_y original magnitude  \\
\hline
&     zb                     & BPZ2.0 most likely redshift  \\
&     zb\_min             &  Lower limit (95p confidence) \\
&     zb\_max             & Upper limit (95p confidence) \\
&     tb                      &  BPZ2.0 most likely spectral type \\
&     Odds                & P(z) integrated within zb - 2$\sigma_{NMAD}$ and zb + 2$\sigma_{NMAD}$\\
&     Chi2                   &  Poorness of BPZ fit: observed vs. model fluxes \\ 
&     Stell\_Mass     &     \texttt{PhotReal} Stellar Mass (log10($M_{\odot}$))  \\
&       LSST\_u, e\_LSST\_u 	          &  LSST\_u  \texttt{PhotReal} magnitude \& uncertainty \\
&       LSST\_g, e\_LSST\_g 	          &  LSST\_g  \texttt{PhotReal} magnitude \& uncertainty \\
&       LSST\_r, e\_LSST\_r	          &  LSST\_r  \texttt{PhotReal} magnitude \& uncertainty \\
&       LSST\_i, e\_LSST\_i 	          &  LSST\_i  \texttt{PhotReal} magnitude \& uncertainty \\
&       LSST\_z, e\_LSST\_z 	          &  LSST\_z  \texttt{PhotReal} magnitude \& uncertainty \\
&       LSST\_y, e\_LSST\_y 	          &  LSST\_y  \texttt{PhotReal} magnitude \& uncertainty \\
\hline
\hline
\end{tabular}
\end{center}
$^*$ as in the original Merson et al. mock catalogue
\end{table*}

\end{document}